\documentclass[11pt,a4paper]{article}

\usepackage{ifthen} 
\newboolean{pdflatex}
\setboolean{pdflatex}{true} 

\newboolean{articletitles}
\setboolean{articletitles}{true} 

\newboolean{uprightparticles}
\setboolean{uprightparticles}{false} 

\newboolean{inbibliography}
\setboolean{inbibliography}{false} 

\usepackage{microtype}
\usepackage{xspace} 
\usepackage{caption} 

\usepackage{graphicx}  
\usepackage{color}
\usepackage{colortbl}
\graphicspath{{./figs/}} 
\usepackage{pdflscape}

\usepackage{siunitx} 
\usepackage{amsmath} 
\usepackage{amssymb}
\usepackage{amsfonts}
\usepackage{upgreek} 

\usepackage{jinstpub}

\usepackage{xspace} 
\usepackage{upgreek}

\def\MagUp {\mbox{\em Mag\kern -0.05em Up}\xspace}

\ifthenelse{\boolean{uprightparticles}}%
{

 \def\PDelta      {\ensuremath{\Delta}\xspace}                 
 \def\PXi      {\ensuremath{\Xi}\xspace}                 
 \def\PLambda      {\ensuremath{\Lambda}\xspace}                 
 \def\PSigma      {\ensuremath{\Sigma}\xspace}                 
 \def\POmega      {\ensuremath{\Omega}\xspace}                 
 \def\PUpsilon      {\ensuremath{\Upsilon}\xspace}                 
 
 \def\PB      {\ensuremath{\mathrm{B}}\xspace}                 
                  
 \def\PD      {\ensuremath{\mathrm{D}}\xspace}

 \def\PK      {\ensuremath{\mathrm{K}}\xspace}

 \def\Pi      {\ensuremath{\mathrm{i}}\xspace}

}
{

 \mathchardef\PDelta="7101
 \mathchardef\PXi="7104
 \mathchardef\PLambda="7103
 \mathchardef\PSigma="7106
 \mathchardef\POmega="710A
 \mathchardef\PUpsilon="7107
                  
 \def\PB      {\ensuremath{B}\xspace}                 
                  
 \def\PD      {\ensuremath{D}\xspace}

 \def\PK      {\ensuremath{K}\xspace}

 \def\Pi      {\ensuremath{i}\xspace}

}

\makeatletter
\ifcase \@ptsize \relax
  \newcommand{\miniscule}{\@setfontsize\miniscule{4}{5}}
\or
  \newcommand{\miniscule}{\@setfontsize\miniscule{5}{6}}
\or
  \newcommand{\miniscule}{\@setfontsize\miniscule{5}{6}}
\fi
\makeatother

\DeclareRobustCommand{\optbar}[1]{\shortstack{{\miniscule (\rule[.5ex]{1.25em}{.18mm})}
  \\ [-.7ex] $#1$}}

  \def\Kbar    {{\kern 0.2em\overline{\kern -0.2em \PK}{}}\xspace}

\def\KorKbar    {\kern 0.18em\optbar{\kern -0.18em K}{}\xspace}

  \def\Dbar    {{\kern 0.2em\overline{\kern -0.2em \PD}{}}\xspace}

\def\DorDbar    {\kern 0.18em\optbar{\kern -0.18em D}{}\xspace}

\def\Bbar    {{\ensuremath{\kern 0.18em\overline{\kern -0.18em \PB}{}}}\xspace}

\def\BorBbar    {\kern 0.18em\optbar{\kern -0.18em B}{}\xspace}

  \def\Y#1S{\ensuremath{\PUpsilon{(#1S)}}\xspace}

\def\Lbar        {{\ensuremath{\kern 0.1em\overline{\kern -0.1em\PLambda}}}\xspace}
\def\LorLbar    {\kern 0.18em\optbar{\kern -0.18em \PLambda}{}\xspace}

\def\AT#1     {\ensuremath{A_{\mathrm{T}}^{#1}}\xspace}           

\def\C#1      {\ensuremath{\mathcal{C}_{#1}}\xspace}                       
\def\Cp#1     {\ensuremath{\mathcal{C}_{#1}^{'}}\xspace}                    
\def\Ceff#1   {\ensuremath{\mathcal{C}_{#1}^{\mathrm{(eff)}}}\xspace}        
\def\Cpeff#1  {\ensuremath{\mathcal{C}_{#1}^{'\mathrm{(eff)}}}\xspace}       
\def\Ope#1    {\ensuremath{\mathcal{O}_{#1}}\xspace}                       
\def\Opep#1   {\ensuremath{\mathcal{O}_{#1}^{'}}\xspace}                    

\newcommand{\unit}[1]{\ensuremath{\mathrm{ \,#1}}\xspace}          

\newcommand{\tev}{\ifthenelse{\boolean{inbibliography}}{\ensuremath{~T\kern -0.05em eV}\xspace}{\ensuremath{\mathrm{\,Te\kern -0.1em V}}}\xspace}
\newcommand{\gev}{\ensuremath{\mathrm{\,Ge\kern -0.1em V}}\xspace}
\newcommand{\mev}{\ensuremath{\mathrm{\,Me\kern -0.1em V}}\xspace}
\newcommand{\kev}{\ensuremath{\mathrm{\,ke\kern -0.1em V}}\xspace}
\newcommand{\ev}{\ensuremath{\mathrm{\,e\kern -0.1em V}}\xspace}
\newcommand{\gevc}{\ensuremath{{\mathrm{\,Ge\kern -0.1em V\!/}c}}\xspace}
\newcommand{\mevc}{\ensuremath{{\mathrm{\,Me\kern -0.1em V\!/}c}}\xspace}
\newcommand{\gevcc}{\ensuremath{{\mathrm{\,Ge\kern -0.1em V\!/}c^2}}\xspace}
\newcommand{\gevgevcccc}{\ensuremath{{\mathrm{\,Ge\kern -0.1em V^2\!/}c^4}}\xspace}
\newcommand{\mevcc}{\ensuremath{{\mathrm{\,Me\kern -0.1em V\!/}c^2}}\xspace}

\def\mum  {\ensuremath{{\,\upmu\mathrm{m}}}\xspace}
\def\muma {\ensuremath{{\,\upmu\mathrm{m}^2}}\xspace}

\def\degc {\ensuremath{^\circ}{C}\xspace}

\def\neutroneq {\ensuremath{\mathrm{ \,n_{eq}}}\xspace}

\def\gsim{{~\raise.15em\hbox{$>$}\kern-.85em
          \lower.35em\hbox{$\sim$}~}\xspace}
\def\lsim{{~\raise.15em\hbox{$<$}\kern-.85em
          \lower.35em\hbox{$\sim$}~}\xspace}

\def\degrees{\ensuremath{^{\circ}}\xspace}

\def\tell1  {TELL1\xspace}
\def\ukl1   {UKL1\xspace}

\def\fig {Figure~}
\def\Fig {Figure~}

\def\sect {Section~}

\def\app {Appendix~}

\def\np {\mbox{{n-on-p}}\xspace}
\def\nn {\mbox{{n-on-n}}\xspace}

\def\maxfluence {{\ensuremath{8 \times 10^{15}~1~\mev \neutroneq~{\mathrm{ \,cm}}^{-2}}\xspace}} 
\def\fluence {\ensuremath{{10^{15}~  1~\mev \neutroneq {\mathrm{ \,cm}}^{-2}}}\xspace}

\DeclareSIUnit\mneq{\cdot\text{1~{MeV~n}$_\text{eq}$\,cm$^{-2}$}\xspace}

\newcommand{\dgc}{\si{\celsius}}

\usepackage{makecell}

\title{Charge collection properties of prototype sensors for the LHCb VELO upgrade}
\author[a,1]{R.~Geertsema\note{Corresponding author},}
\author[a]{K.~Akiba,}
\author[a]{M.~van Beuzekom,}
\author[b]{E.~Buchanan,}
\author[c]{C.~Burr,}
\author[c,d]{W. Byczynski,}
\author[c]{P.~Collins,}
\author[a,e]{E.~Dall'Occo,}
\author[c,f]{T.~Evans,}
\author[g]{V.~Franco Lima,}
\author[a]{K.~Heijhoff,}
\author[h]{P.~Kopciewicz,}
\author[i]{F.~Marinho,}
\author[b]{E.~Price,}
\author[h]{B.~Rachwal,}
\author[b]{S.~Richards,}
\author[b]{D.~Saunders,}
\author[c]{H.~Schindler,}
\author[a]{H.~Snoek,}
\author[h]{T.~Szumlak,}
\author[a,j]{P.~Tsopelas,}
\author[b]{J.~Velthuis,}
\author[k]{and M.R.J.~Williams}

\affiliation[a]{Nikhef, Science Park 105, 1098 XG Amsterdam, the Netherlands}
\affiliation[b]{University of Bristol, Beacon House,\\Queens Road, BS8 1QU, Bristol, United Kingdom}
\affiliation[c]{CERN, 1211 Geneve, Switzerland}
\affiliation[d]{Tadeusz Kosciuszko Cracow University of Technology, Cracow, Poland}
\affiliation[e]{Now at TU Dortmund,\\Otto-Hahn-Straße 4, 44227 Dortmund, Germany}
\affiliation[f]{University of Oxford, Particle Physics Department,\\Denys Wilkinson Bldg., Keble Road, Oxford OX1 3RH, United Kingdom}
\affiliation[g]{Oliver Lodge Laboratory, University of Liverpool, \\Liverpool, L69 7ZE, United Kingdom}
\affiliation[h]{AGH University of Science and Technology, Faculty of Physics and \\Applied Computer Science, Krak\'ow, Poland}
\affiliation[i]{Federal University of S\~ao Carlos,\\Rodovia Anhanguera, km 174, 13604-900 Aranas, Brazil}
\affiliation[j]{Now at Spectricon, Science and Technology Park of Crete, Heraklion, Greece}
\affiliation[k]{School of Physics and Astronomy, University of Edinburgh, \\Edinburgh, United Kingdom }

\emailAdd{r.geertsema@nikhef.nl}

\abstract{

An extensive sensor testing campaign is presented, dedicated to measuring the 
charge collection properties of prototype candidates for the 
Vertex Locator (VELO) detector for the upgraded LHCb experiment.
The charge collection is measured with sensors exposed to fluences of up to \maxfluence, as well as with
nonirradiated prototypes. The results are discussed, 
including the influence of different levels of irradiation and bias voltage on the charge collection properties.
Charge multiplication is observed on some sensors that were nonuniformly irradiated with 24\gev protons, to the
highest fluence levels. An analysis of the charge collection near the guard ring region
is also presented, revealing  significant differences between the sensor prototypes. 
All tested sensor variants succeed in collecting the minimum required charge of 6000 electrons after the exposure to the maximum fluence.

}


\begin{document}
\maketitle


%


\section{Introduction}

The Vertex Locator (VELO) \cite{Aaij_2014} is the silicon vertex detector surrounding the interaction region of the LHCb experiment \cite{Collaboration_2008}. 
Throughout LHC Runs 1 and 2, the VELO was based on silicon strip sensors. The VELO will be upgraded to a hybrid pixel system for Runs 3 and 4. The detector is designed to withstand a maximum fluence of \maxfluence, corresponding to the amount expected after an integrated luminosity of 50\unit{fb^{-1}}\cite{TDR} over the 8-year lifetime of the upgraded experiment. The charge collection properties of the prototype sensors prior to, and after irradiation up to maximum fluence are reported in this paper.

An extensive R\&D programme was launched to obtain sensors able to meet the challenging requirements for radiation hardness of the VELO upgrade. 
One of the crucial sensor requirements is that the amount of charge collected for a high energy charged particle is more than 6000\unit{e^-} after the detector is exposed to the maximum fluence. 
This ensures that the signal is sufficiently above the expected highest threshold of 1000\unit{e^-}, even if the charge is shared amongst several pixels.
Furthermore, it was essential to evaluate the performance at the edge of the sensor, since this is the region closest to the interaction point and hence with the highest occupancy during the operations of the detector.

The prototype sensors have been characterised using readily available Timepix3 ASICs\cite{Poikela_2014}. The Timepix3 ASIC has a data-driven readout which provides a 10 bit Time-over-Threshold (ToT) measurement, and a timestamp with $\sim1.56$\,ns precision.

In this paper, the performance of several different prototypes is discussed, comparing results of irradiated and nonirradiated samples. The different sensor designs are described in  \sect\ref{sec:Prototype}. The experimental setup is outlined in \sect\ref{sec:telescope}, with a detailed description of the charge calibration method in \sect\ref{Sec:Calibration}. The irradiation programme is detailed in \sect\ref{Sec:Irradiation}. The charge collection results for different bias voltages are discussed in \sect\ref{sec:BiasStudies}. During the evaluation programme, radiation induced charge multiplication was observed and results are described in \sect\ref{Sec:ChargeMultiplication}. The edge performance is presented in \sect\ref{sec:edge}, followed by the conclusion in \sect\ref{sec:conclusion}.

\section{Sensor prototype designs}
\label{sec:Prototype}

A VELO upgrade assembly consists of a single sensor bump-bonded to three ASICs, and is referred to as a triple-chip assembly. However, for the charge collection studies, single ASIC assemblies have also been used, since it makes the irradiation and testing in the lab or in the beam more affordable.
The size of the sensors for the VELO upgrade is approximately $42.5\times14.1$ mm$^2$,
consisting of three groups of $256\times256$ pixels. The pixels have a 55\mum\ pitch, except for those in the interchip region, which are elongated in order to have complete coverage between the ASICs. The elongated pixels are either 110 or 137.5\mum depending on the round of production. 

\begin{figure}[tb]
  \centering 
  \includegraphics[width=0.95\textwidth]{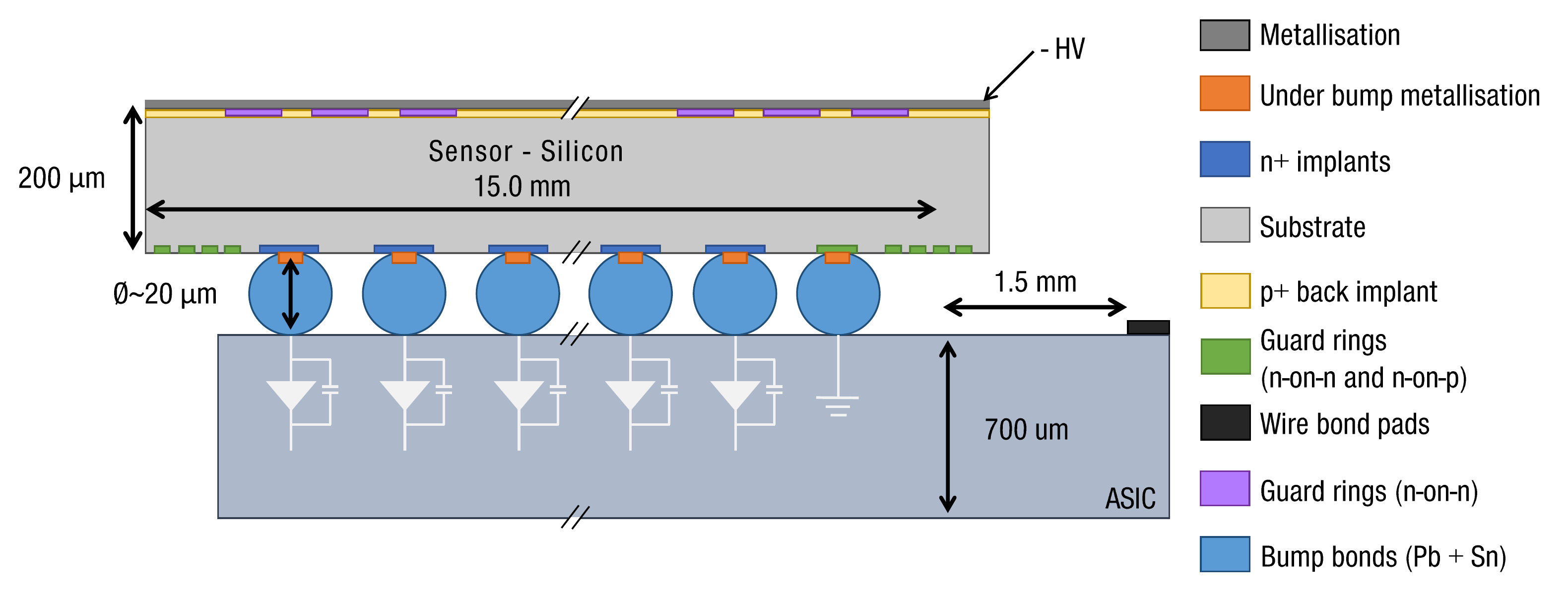}
  \caption{ A cross section of the sensor connected to the ASIC. The different implants are indicated with different colors. Both the design of the \np and \nn sensors are shown in this diagram, since the only difference is the guard rings on the back of the sensor.}
  \label{fig:sensorcrosssec}
\end{figure}

For the first round of prototyping, sensors have been produced by two vendors, Hamamatsu Photonics K. K. (HPK)\footnote{Hamamatsu Photonics K. K., 325-6, Sunayama-cho, Naka-ku, Hamamatsu City, Shizuoka, 430-8587, Japan} and Micron Semiconductor Ltd\footnote{Micron Semiconductor Ltd, 1 Royal Buildings, Marlborough Road, Lancing BN158UN, United Kingdom
}. 
The HPK sensors have been produced with n$^+$-type implants separated by p$^+$-stop implants on a $200\pm 20 $\mum\ thick float-zone p-doped silicon substrate   
(resistivity of 3-8 k$\Omega$ cm). The back of the sensor consists of a thin p$^+$-doped layer and is fully metallised. 
Two different guard ring designs with distances of 450 and 600\mum from the edge of the sensor to the edge of pixel matrix have been evaluated. This distance is often referred to as Pixel-To-Edge (PTE). 
The pixel implants are either 35 or 39\mum squares with rounded corners. The Micron prototypes have been produced with 36\mum\ wide n$^+$-type implants with rounded corners and p$^+$-spray isolation. Two different types of substrates have been used: 200\mum\ p-type ($>$ 5 k$\Omega$ cm) and 150\mum\ n-type ($>$ 1.5 k$\Omega$ cm). The latter is double sided processed with guard rings on the backside of the sensor partially implanted underneath the edge pixels (see \fig\ref{fig:sensorcrosssec}). The back of these sensors consist of a thin p$^+$-doped layer and the back for the p-type substrate is fully metallised, while for the n-type substrate the back is metallised in a grid structure.
For these sensors there are two PTE variants with corresponding distances of 250\mum and 450\mum. The n$^+$-p-p$^+$ (n$^+$-n-p$^+$) sensors are from now on referred to as \np (\nn) sensors.

The operational voltage needed to yield a signal of at least 6000\unit{e^-} after the highest fluence is expected to be \SI{1000}{\volt}~\cite{TDR}.
For \np sensors
the operation above a certain voltage can generate an electrical discharge 
between the sensor and the ASIC if the assembly is placed in a gaseous environment. The \nn sensor design does not suffer from electrical discharge at voltages below \SI{1000}{\volt}
since these sensors are processed on both sides allowing for guard rings on each side of the sensor. This lowers the potential on the side closer to the ASIC.
In order to prevent discharges in \np sensors, these sensors either received a C-type parylene coating\footnote{Comelec SA, La Chaux-de-Fonds, Switzerland}
prior to irradiation or are operated in vacuum.

\section{Experimental setup and calibration}
\label{sec:telescope}

\subsection{The Timepix3 telescope at CERN SPS}
An extensive testbeam programme has been carried out at the SPS H8
beamline at CERN to characterise the sensors at high rate.  
The beam is a mixed charged hadron beam 
($\sim 67\%$ protons, $\sim30\%$ pions) at 180\gevc.

The Timepix3 telescope~\cite{tpx3telescope} is a high rate, data-driven beam telescope, composed of two arms of four planes each.
Each plane is equipped with a 300\mum p-on-n silicon sensor 
that is bump bonded to a Timepix3 ASIC. 
The centre of the telescope is reserved for the Device Under Test (DUT). 
The DUT area is equipped with remotely controlled motion stages able 
to translate in $x$ and $y$ directions (orthogonal to the beam axis) 
and rotate about the $y$ axis. 
A vacuum box can also be installed on the central stage  to facilitate 
testing of irradiated devices at high voltage. The cooling block connected to the ASIC can be cooled down 
to temperatures of  $ -35\  \dgc$, which keeps the sensors below $ -20\  \dgc$. The temperature of the cold box is measured with a Pt100, which is converted to the corresponding sensor temperature using a calibration curve. This calibration curve is determined before the testbeam in a dedicated lab setup. In this setup a Pt100 was glued to the sensor itself and another Pt100 was glued to the cold box in order to measure the sensor temperature for different cold box temperatures.
The pointing resolution at the DUT position is about $1.6\mum$, 
enabling intrapixel studies of the DUT.
The typical temporal resolution on a track using only the Timepix3 timestamps 
is about $350$~ps.

\subsection{Front-end calibration}
\label{Sec:Calibration}

The recorded Time-over-Threshold (ToT) values are converted to a corresponding number of electrons via a charge calibration procedure. 
The measured ToT for a given injected charge varies from pixel to pixel 
due to variations in the discharge current across the pixel matrix. 
The relation between ToT and charge for each pixel can be determined by 
injecting testpulses with known charge into the preamplifier. The injected charge depends on the voltage difference
of two internal voltage DACs whose value can be measured externally, 
and on the capacitance of the test capacitor which varies only slightly across the ASIC.

A calibration curve is determined separately for each pixel \cite{JAKUBEK2008155}.
The test pulse calibration method was validated for one of the assemblies 
by comparing it to the characteristic lines from an $^{241}$Am source and agreed to within 4\% \cite{VicenteBarretoPinto:2134709}, and was also  cross-checked using X-rays with several
energies from the LNLS synchrotron in Brazil.
 

\section{Irradiation programme}
\label{Sec:Irradiation}

The fluence that the VELO upgrade will receive is estimated 
using simulations of proton collisions at 14 TeV~\cite{TDR}. 
The hadronic collisions produce a nonuniform fluence exposure over each  sensor. 
The maximum integrated fluence expected at the most irradiated
sensor position is \maxfluence. 
The difference in fluence over a single sensor due to
this nonuniform irradiation can be as large as a factor 120.

Three different types of irradiation are used in the studies presented in this paper. 
Sensors received uniform neutron irradiation at the TRIGA MARK II reactor at JSI~\cite{JSI} in Ljubliana, Slovenia, of half, and of maximum fluence.
The IRRAD facility at CERN~\cite{Gkotse_2237333} provides a 24~\gev proton beam that is approximately normally distributed, hence yielding a nonuniform fluence profile.
Forschungszentrum Karlsruhe Cyclotron at Karlsruhe (KIT) 
can provide a 23\,\mev\ proton beam with a small cross section which can be scanned over the sensor in order 
to approximate the nonuniform radiation profile of the upgraded VELO. 
The fluences provided by JSI, IRRAD, and KIT are known with an accuracy of $\pm10\%$, $\pm10\%$, and $\pm20\%$, respectively. Table~\ref{tab:IrradFacilities} shows a summary of the properties of the three facilities.

\begin{table}
\centering
\caption{Characteristics of the  facilities used in the irradiation programme.}
  \begin{tabular}{lccccc}
  \hline
  Facility   & Particles             & Cooling & Scanning  &   \makecell{Intensity \\  {[$10^{12}$\,s$^{-1}$\,cm$^{-2}$]} } & \makecell{Hardness \\ {Factor}} \\
  \hline 
  Karlsruhe  & 23\,\mev\ protons   & yes     & yes       & 25                                          &$2.20\pm0.40$ \cite{Allport_2019} \\
  IRRAD      & 24\,\gev\ protons   & yes     & no        & 0.02                                        & $0.62\pm0.04$ \cite{Allport_2019} \\
  Ljubljana  & reactor  neutrons  & no      & no        & 3.05                                        &$0.90\pm0.03$ \cite{Kramberger:1390490} \\
  \hline
  \end{tabular}
    
    \label{tab:IrradFacilities}
\end{table}

\subsection{IRRAD profile}

\begin{figure}[tb]
  \centering 
  \includegraphics[width=0.47\textwidth]{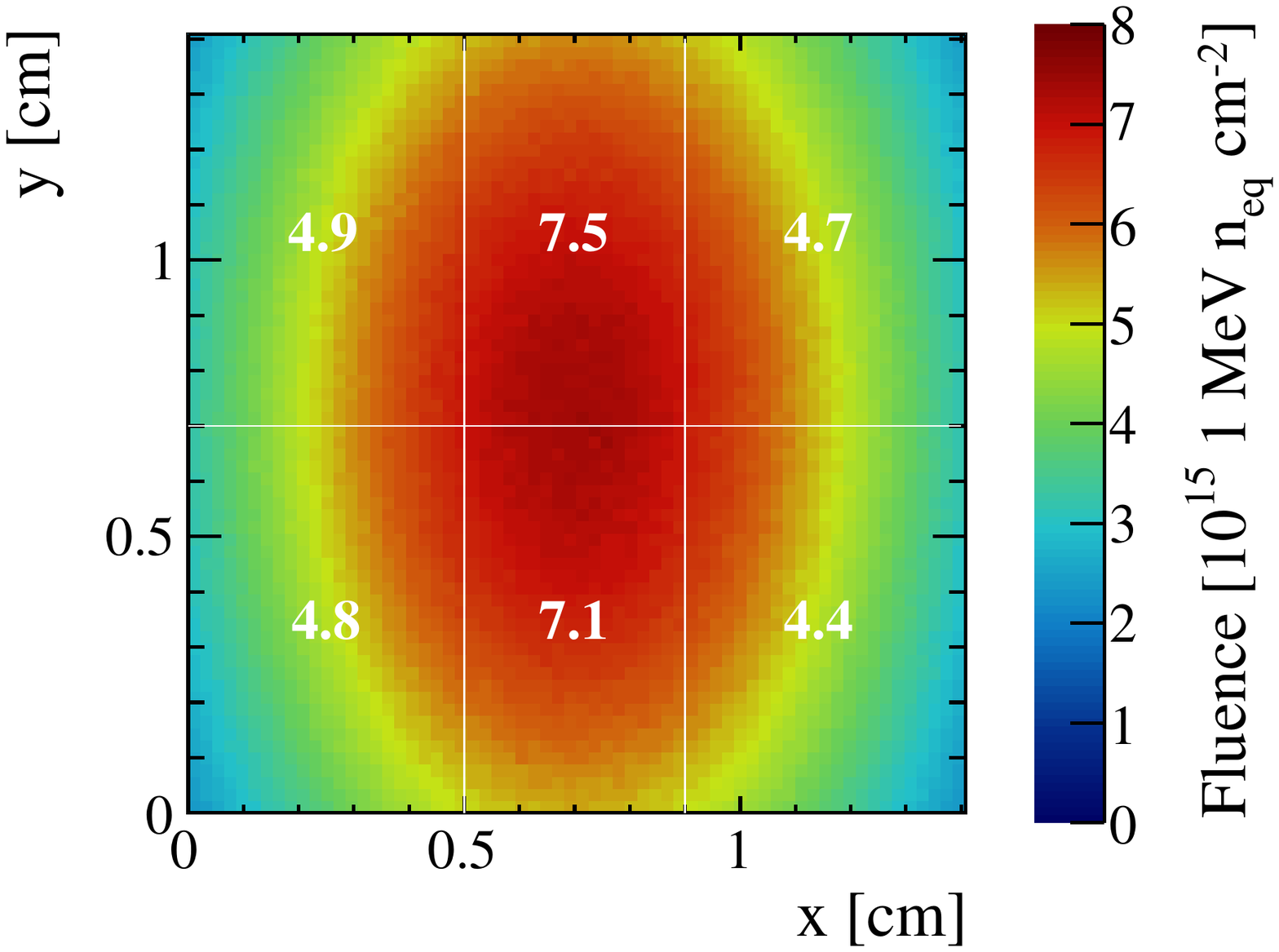}
  \caption{The reconstructed fluence profile from irradiation at IRRAD. The dosimetry results, converted to neutron equivalent fluence, for the six regions are indicated by the white numbers ($\times$\fluence).}
  \label{Fig:IRRADprofile}
\end{figure}

After irradiation, the residual activation of the assemblies is obtained by measuring the distribution of hits in the sensor when not exposed to the beam.
The hits are caused by the radioactivity induced in the assembly and thus their rate is proportional to the fluence. 
The radiation profile can then be  determined from the activation map, modelled using a two-dimensional Gaussian distribution~\cite{DallOccoThesis}. 
The activation profile is then normalised to match the fluence measured by an aluminium foil that was placed alongside the assemblies during irradiation.
The foil was later subdivided in six regions, examined for activation through $\gamma$-ray spectroscopy.
The reconstructed fluence profile of a single-chip assembly is shown in \fig\ref{Fig:IRRADprofile}.
The corresponding fluence measured with the aluminium foil is overlaid for each region, where the measured proton fluence has been converted in neutron equivalent unit using a hardness factor of $0.62\pm0.04$~\cite{Allport_2019}. 
The reconstructed fluence profile is found to be consistent with the measured charge collection profile.

\section{Results with beam}
\label{sec:Results}

The prototype sensors were characterised using the Timepix3 telescope in beam tests using the SPS at CERN. 
Only DUT clusters associated to a telescope track are selected in order to mitigate background hits from internal radioactivity and noise.
Clusters are matched to a track if they are within a space (time) region of 110$\times$110\muma (\SI{100}{\nano s}) from the value predicted by the telescope.
The cluster size is limited to four pixels since larger clusters are usually 
due to nuclear interactions (with a much larger energy deposition) or $\delta$-rays.

A fit is performed to the cluster charge distribution, which can be described by a convolution of Landau and Gaussian distributions, in order to determine the Most Probable Value (MPV) of the Landau.
The standard deviation of the MPV
distribution over the chip is quoted as the uncertainty on the MPV, and is determined for this assembly to be around $350$\unit{e^-}, corresponding to about $2\%$ of the MPV, which is much
larger than the statistical estimate from the fit.
This uncertainty is determined with a single nonirradiated 
200\mum $\,$\np assembly and, since it is dominated by ASIC effects, it is considered representative of all other prototypes.
The uncertainty on the absolute calibration is determined to be $4\%$, as discussed in \sect\ref{Sec:Calibration}.

This section is organised as follows. In \sect\ref{sec:BiasStudies} the charge distribution is studied for both nonirradiated and uniformly irradiated sensors operated at different bias voltages and placed perpendicular to the beam. It is followed by \sect\ref{Sec:ChargeMultiplication} in which the charge distribution as a function of fluence for nonuniformly irradiated sensors also placed perpendicular to the beam is discussed. Finally, the performance at the edge of the sensors is discussed in \sect\ref{sec:edge}, in which different angles of the sensor with respect to the beam are used.
For figures in which devices of different types are shown together, a colour code is used: green for HPK \np, blue for Micron \np, and purple for Micron \nn. An overview of the assemblies is given in \app\ref{sec:appendixAssemblies}.

\subsection{Bias Studies}
\label{sec:BiasStudies}

\subsubsection*{Nonirradiated assemblies}
Assemblies were tested in the SPS beam at the same time as the irradiation campaign, thus not all of the assemblies could be studied prior to being irradiated; therefore, a representative subset of assemblies was tested prior to being irradiated in order to make comparisons with the results post irradiation.

\begin{figure}
\centering
\begin{minipage}{.47\textwidth}
\centering
\includegraphics[width=1\linewidth]{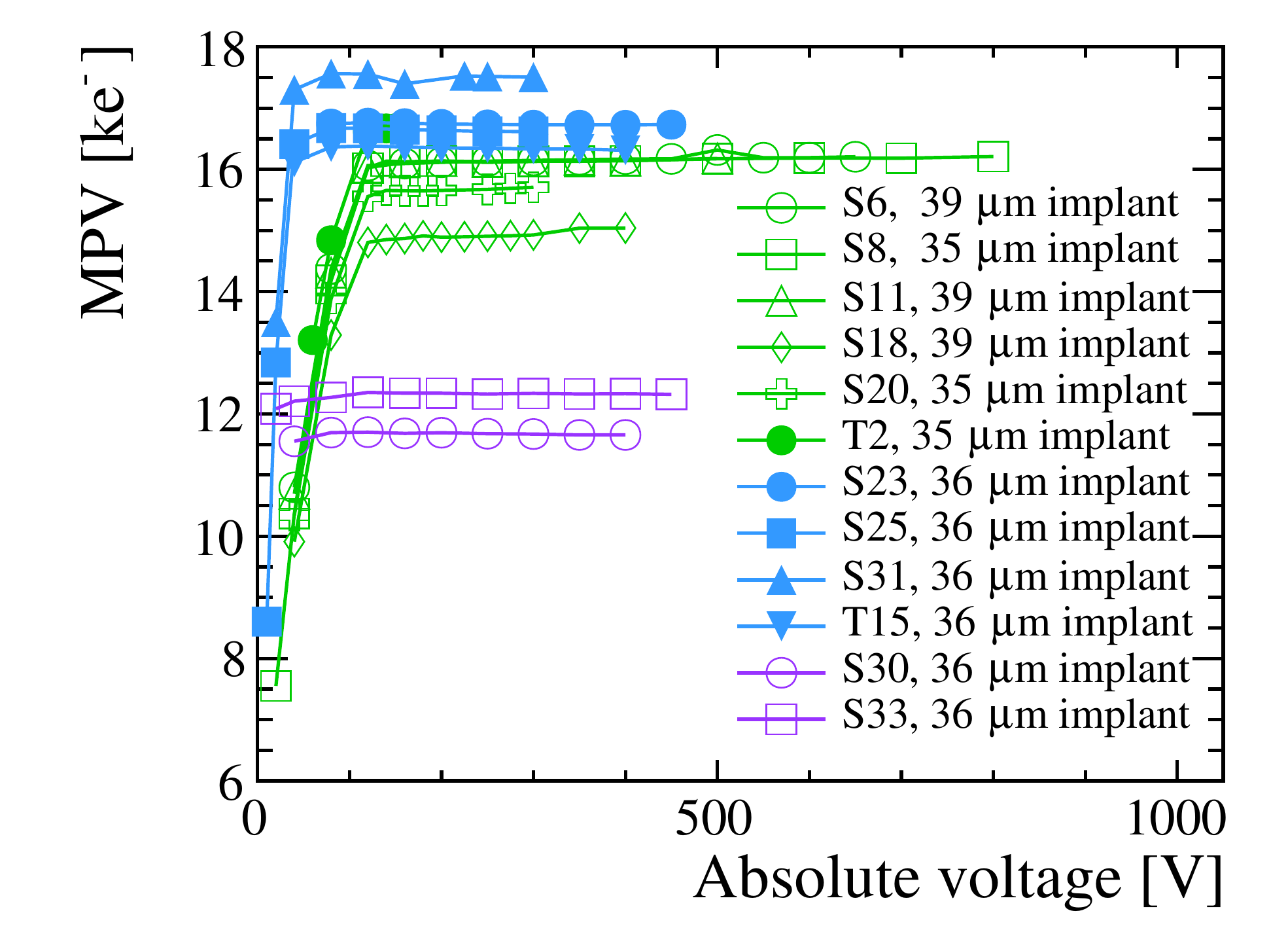}
\end{minipage}\qquad
\begin{minipage}{.47\textwidth}
\centering
\includegraphics[width=1\linewidth]{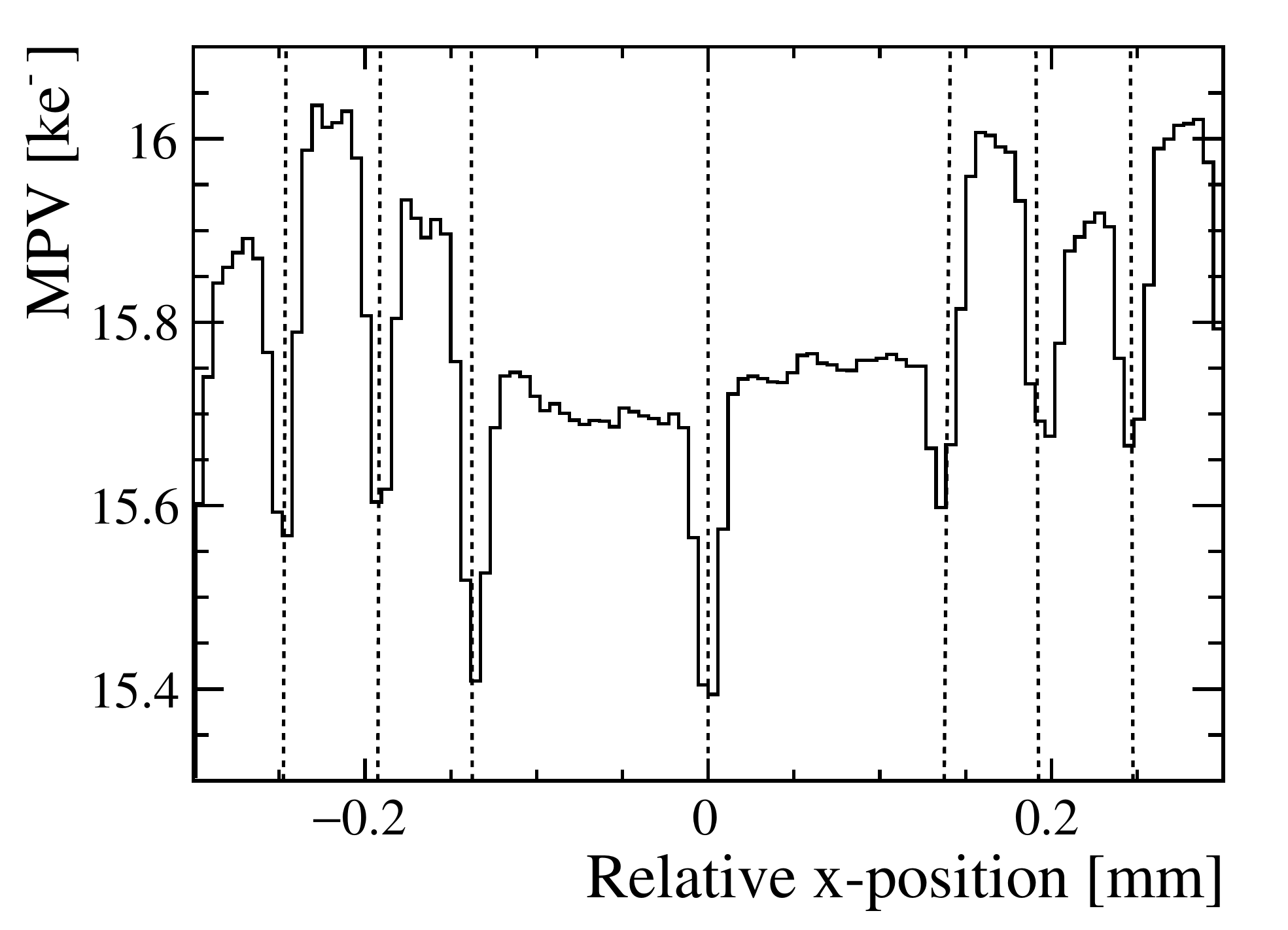}
\end{minipage}

\begin{minipage}[t]{.47\textwidth}
\centering
\captionsetup{width=.99\linewidth}
\captionof{figure}{MPV of the cluster charge (for clusters associated to a telescope track) as function of bias voltage 
           for nonirradiated assemblies,
           measured in beam tests. 
           The uncertainties are not indicated in the plot, see the text for a detailed explanation. 
           }
\label{Fig:MPVvsBiasTestbeamBeforeIrradiation}
\end{minipage}\qquad
\begin{minipage}[t]{.47\textwidth}
\centering
\captionsetup{width=.99\linewidth}
\caption{The MPV for different positions around the inter-chip 
  region between two ASIC on T23 before irradiation. The middle of the
  interchip region is indicated as the zero point on the x-axis. The vertical dashed lines indicate the pixel column borders. }
\label{fig:interChipMPV}
\end{minipage}
\end{figure}

\Fig\ref{Fig:MPVvsBiasTestbeamBeforeIrradiation} shows the MPV as a
function of bias voltage for several nonirradiated assemblies. 
For the HPK sensors, indicated by the green curves, the MPV of the collected charge saturates 
between 100 and \SI{110}{\volt} and shows no dependence on the implant size.
The MPV of the Micron sensors, indicated by the blue curves 
(\np) and purple curves (\nn), reaches a plateau at 
less than $40$\,V. The difference in MPV for the Micron sensors 
is due the different thicknesses of the two types, 200\mum and 150\mum for \np and \nn sensors, respectively.
The small differences in MPV between sensors of the same type are due to wafer-to-wafer variations in thickness up to 10\%, and variation in the testpulse capacitance due to process variations of about 5\% (one sigma).

The main difference other than the area between the single-chip and triple-chip assemblies is the
presence of elongated pixels located in the region in between the chips. 
It is important that the response of the inter-chip region is similar to the rest of the sensor. This region is investigated for 
the HPK \np triple-chip assembly T23 before irradiation. 
The MPV of the interchip region of HPK triple-chip assembly T23 is show in \fig\ref{fig:interChipMPV}, where the middle between chips is defined as zero.
Three characteristics can be observed: the dips between columns; the variation from column-to-column (the narrow flat regions); and the slightly lower MPV of the interchip region (wider flat regions).
The dips between columns are due to threshold effects, and the variation from column-to-column is due to the charge calibration.
The MPV is slightly lower for the two elongated columns compared to the other columns in the figure. However, this difference still falls within the total variation of the MPV over the columns, which is found to be $300$\unit{e^-} using the same method as for the total variation over the pixel matrix. Therefore, it is concluded that the interchip region collects all charge.

\begin{figure}[tb]
  \centering 
  \includegraphics[width=0.47\textwidth]{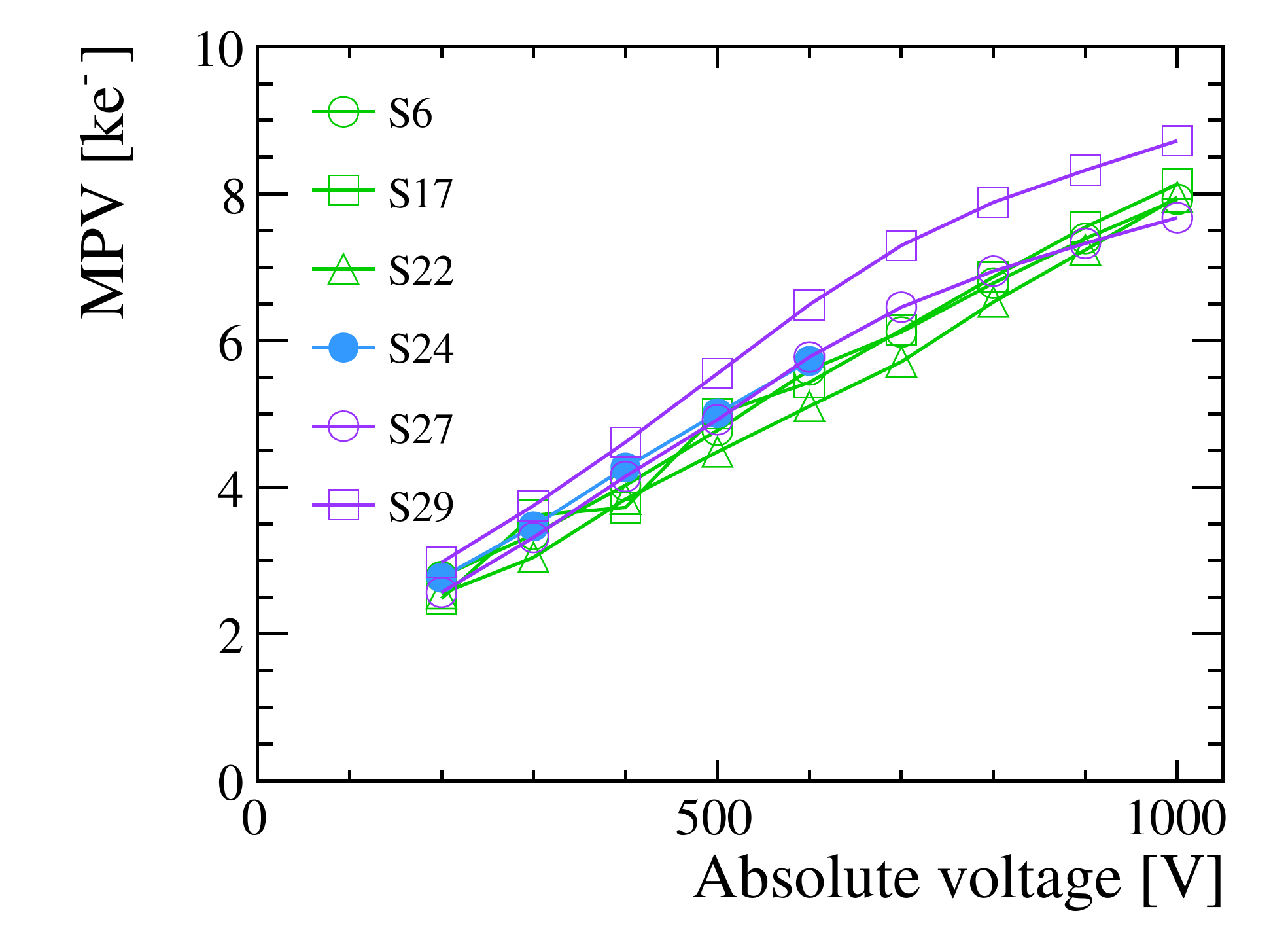}
  \caption{MPV as function of bias voltage after uniform irradiation 
           at JSI Ljubljana to \maxfluence, without additional controlled annealing.  The uncertainties are not indicated in the plot, see the text for a detailed explanation.}
  \label{Fig:MPVvsBiasTestbeamJSI}
\end{figure}

\subsubsection*{Irradiated assemblies}

The MPV as a function of bias voltage after irradiation to full fluence at JSI is shown in \fig\ref{Fig:MPVvsBiasTestbeamJSI}.
All of the assemblies follow the same trend, with the MPV increasing 
linearly from about $2500$\,\unit{e^{-}} at \SI{200}{\volt} to about $8000$\unit{e^-} at \SI{1000}{\volt}.  
The leakage current at bias voltages higher than 600~V for S24 was larger than the leakage current compensation in the ASIC, resulting in changes in the charge calibration per pixel.
Hence these points are excluded from this figure.

\begin{figure}
\centering
\begin{minipage}{.47\textwidth}
\centering
\includegraphics[width=1\linewidth]{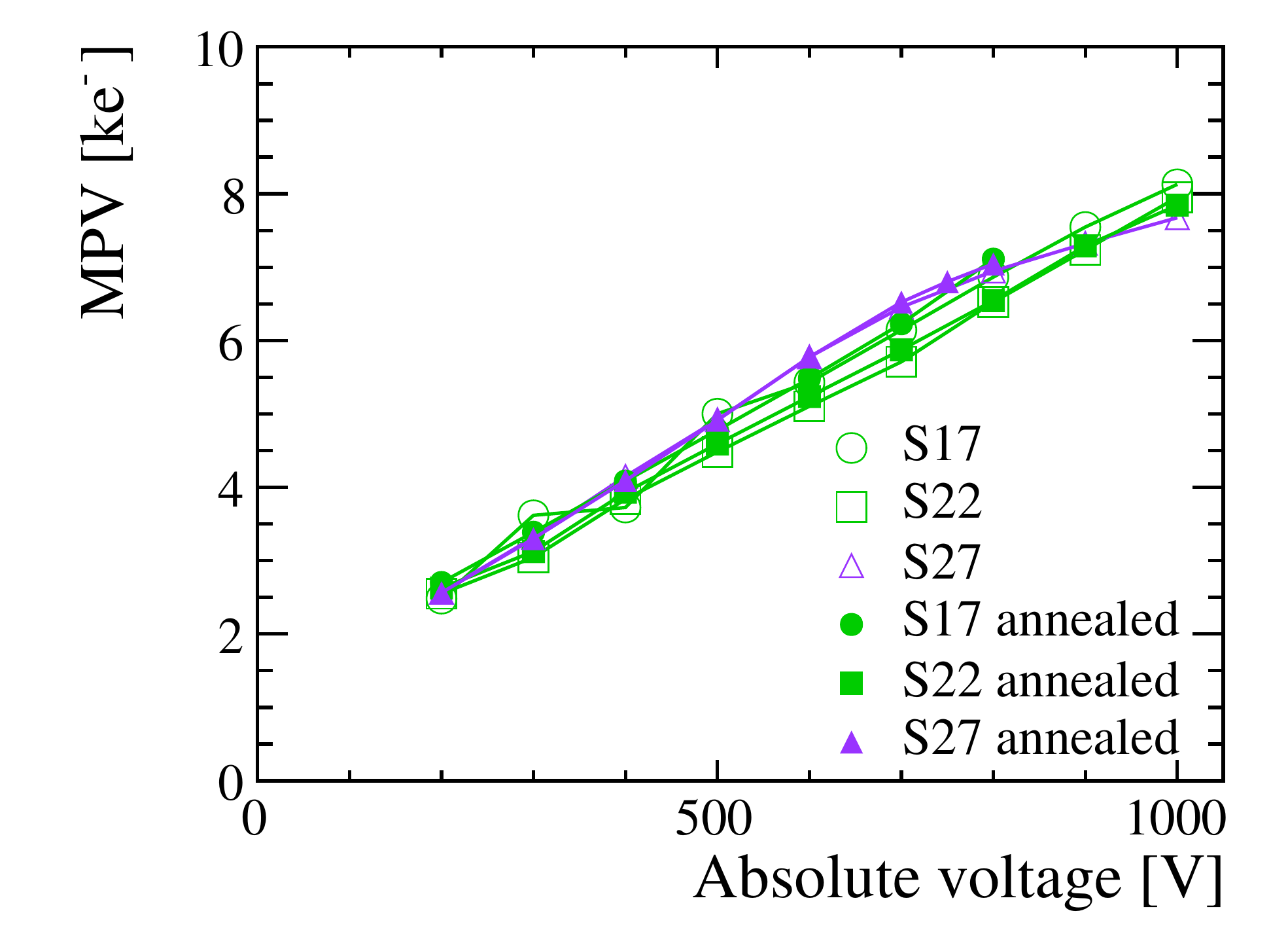}
\end{minipage}\qquad
\begin{minipage}{.47\textwidth}
\centering
\includegraphics[width=1\linewidth]{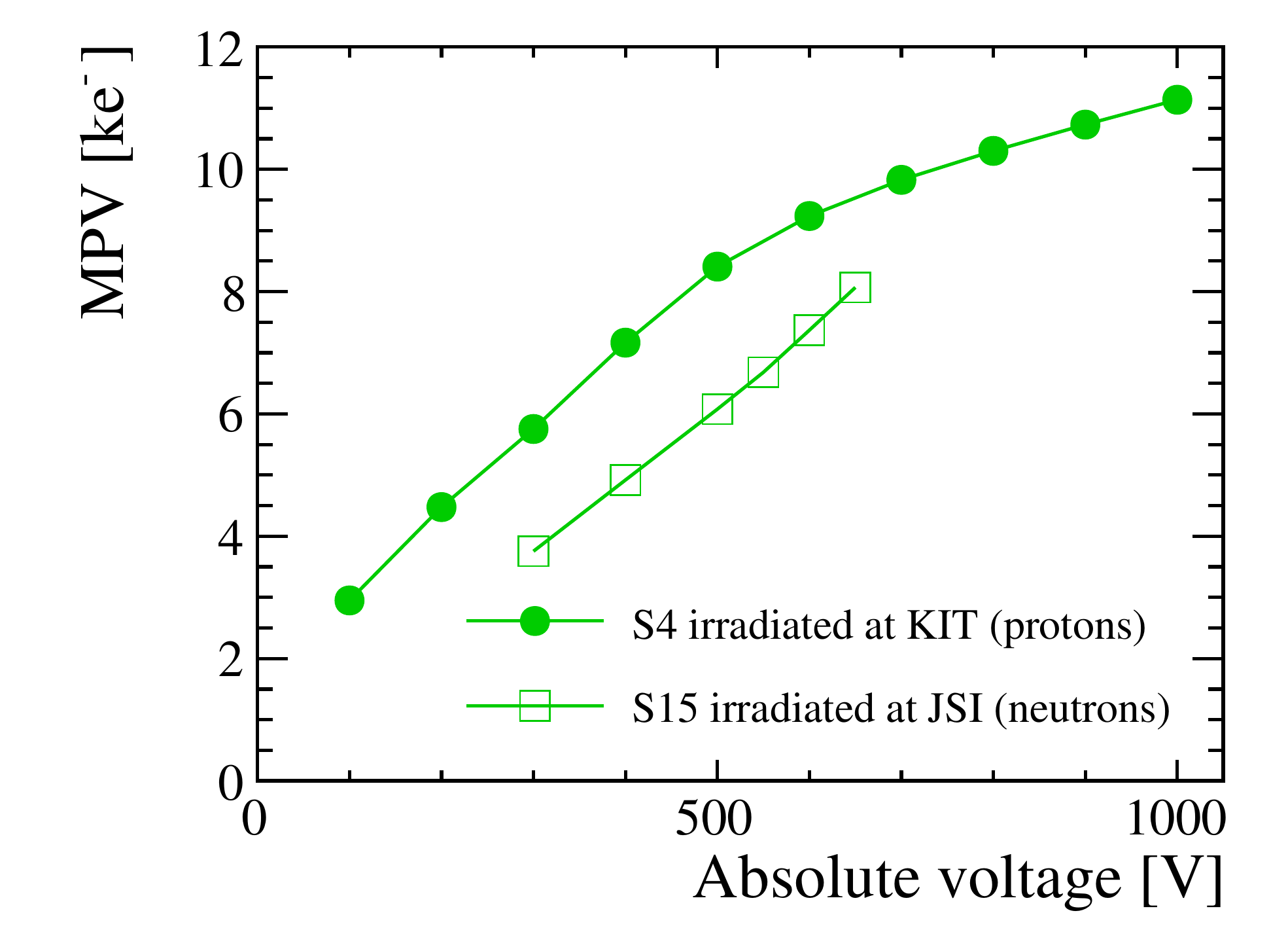}
\end{minipage}


\begin{minipage}[t]{.47\textwidth}
\centering
\captionsetup{width=.99\linewidth}
\captionof{figure}{MPV as function of bias voltage after uniform irradiation 
           at JSI to \maxfluence, 
           before and after annealing for 80\,min. at 60\degc.}
\label{Fig:MPVvsBiasTestbeamJSIAnnealing}
\end{minipage}\qquad
\begin{minipage}[t]{.47\textwidth}
\centering
\captionsetup{width=.99\linewidth}
\caption{MPV as function of bias voltage after uniform irradiation 
           to $4\times$\fluence,
           without additional controlled annealing.}
\label{Fig:MPVvsBiasTestbeamHalfFluence}
\end{minipage}
\end{figure}

The measurements shown in \fig\ref{Fig:MPVvsBiasTestbeamJSI} were performed 
before any controlled annealing. A subset of the assemblies were 
tested again with beam after having been annealed for 80 minutes at 60~\degc. 
As can be seen in \fig\ref{Fig:MPVvsBiasTestbeamJSIAnnealing}, the results before
and after annealing are in agreement. 
The leakage current at bias voltage of 900~V for S17 after annealing was larger than the leakage current compensation in the ASIC, resulting in changes in the charge calibration per pixel. The sensors were kept at room temperature for 11 days after irradiation, after which they were cooled at -15~\degc, beside the aforementioned additional controlled annealing.

Several assemblies have been irradiated with either reactor neutrons or protons. 
As described in reference~\cite{CINDRO200960}, the damage to silicon from neutron irradiation 
is different to the damage from proton irradiation. 
\Fig\ref{Fig:MPVvsBiasTestbeamHalfFluence} shows the MPV as a function of bias voltage
for two HPK single-chip assemblies after uniform irradiation to a fluence of 
$4\times$\fluence at KIT (assembly S4 - proton irradiation) 
and JSI (assembly S15 - neutron irradiation). 
The curve for S15 does not extend up to \SI{1000}{\volt} because the assembly sparked, as it was not parylene coated and thus measurements were performed only up to \SI{675}{\volt} in a CO$_{2}$ atmosphere.

\subsection{Charge multiplication}
\label{Sec:ChargeMultiplication}

As reported in several previous studies \cite{MANDIC2009263, LANGE201049, MIKUZ2011S50}, charge multiplication in planar silicon sensors is expected at fluences above 
{\ensuremath{5 \times 10^{15}~1~\mev  \neutroneq~{\mathrm{ \,cm}}^{-2}}\xspace} 
but typically in considerably larger pitches than presented here.

\begin{figure}
  \centering
  \includegraphics[width=0.48\textwidth]{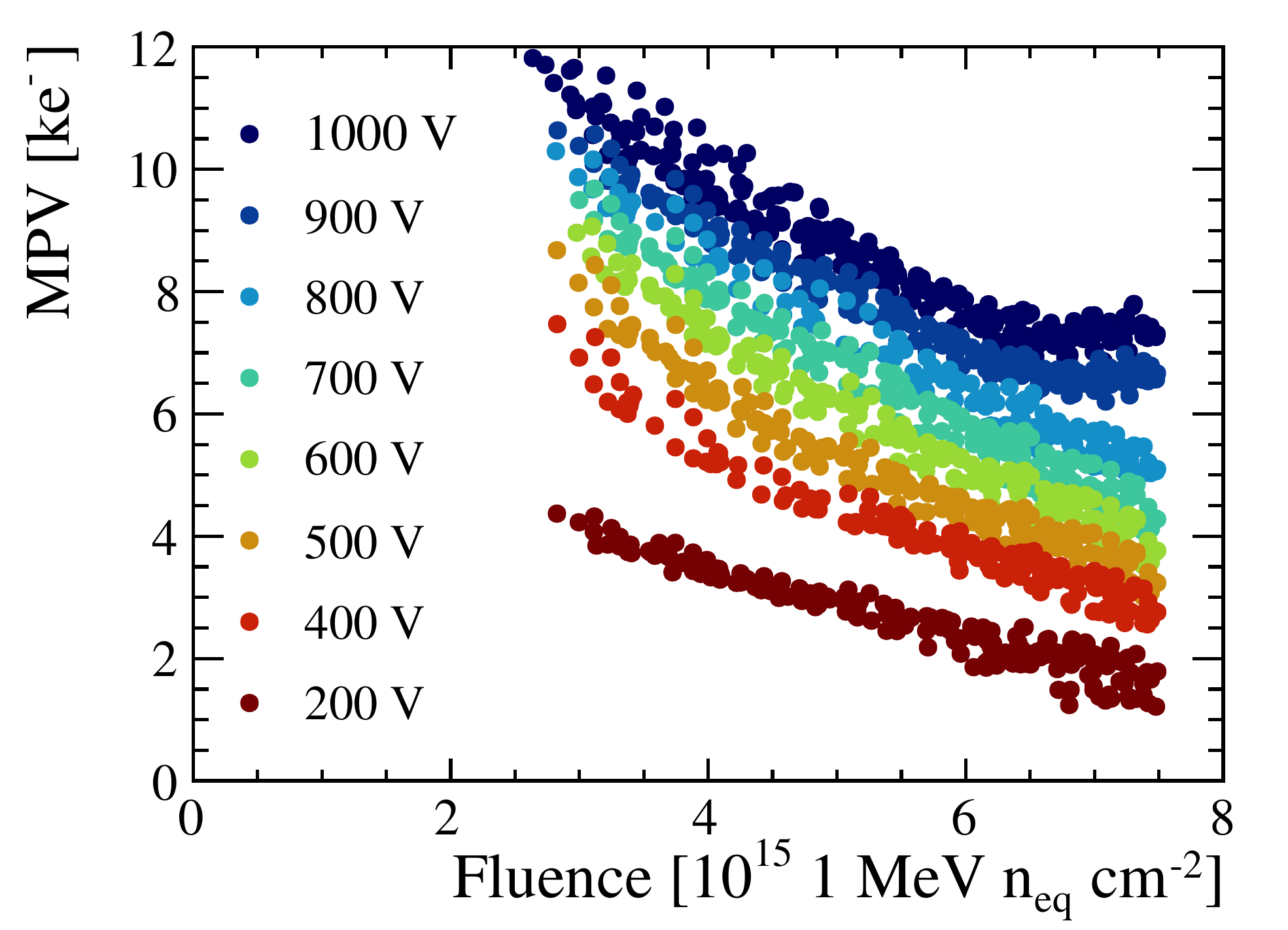}
  \includegraphics[width=0.48\textwidth]{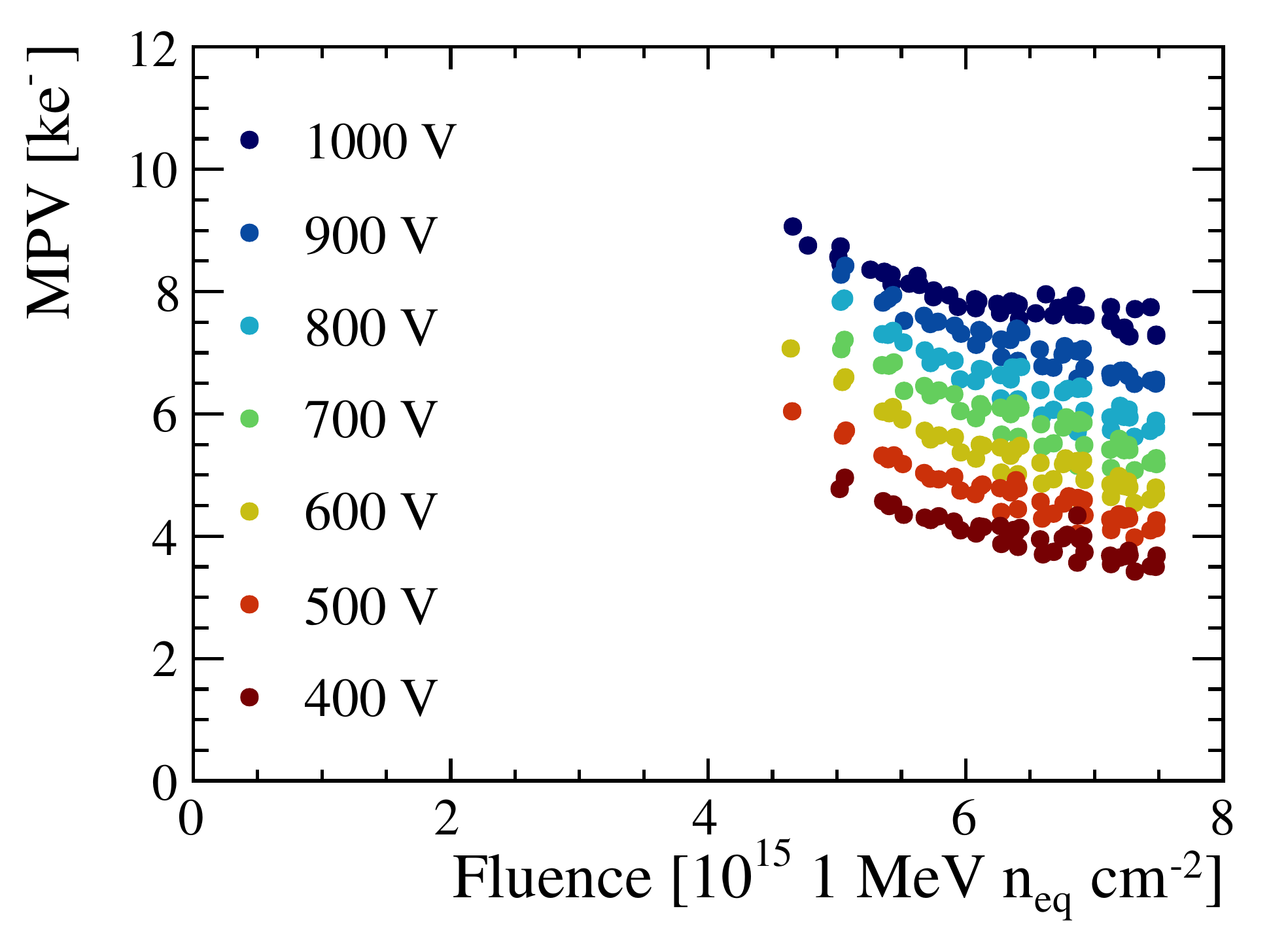}
  \includegraphics[width=0.48\textwidth]{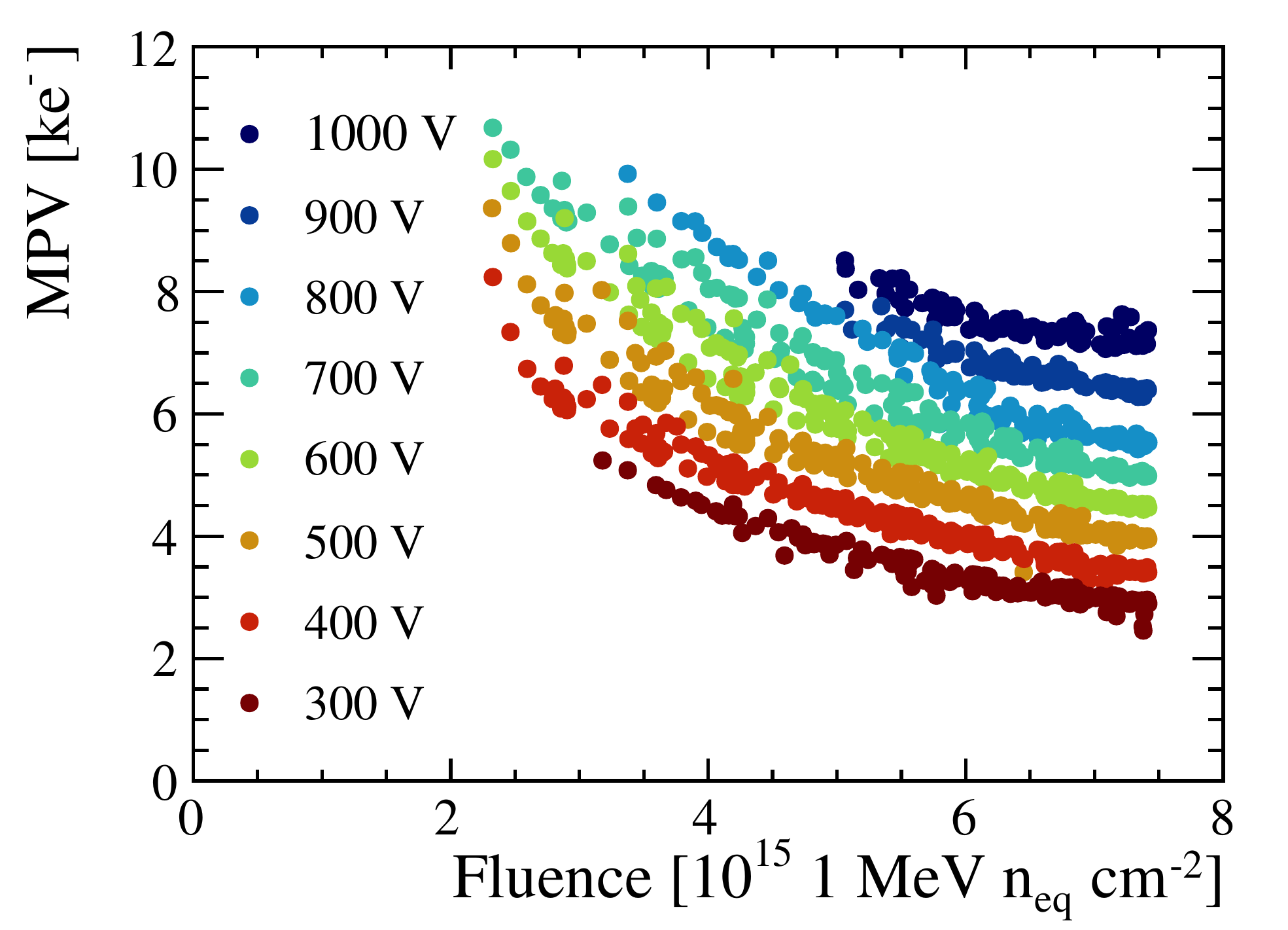}
  \includegraphics[width=0.48\textwidth]{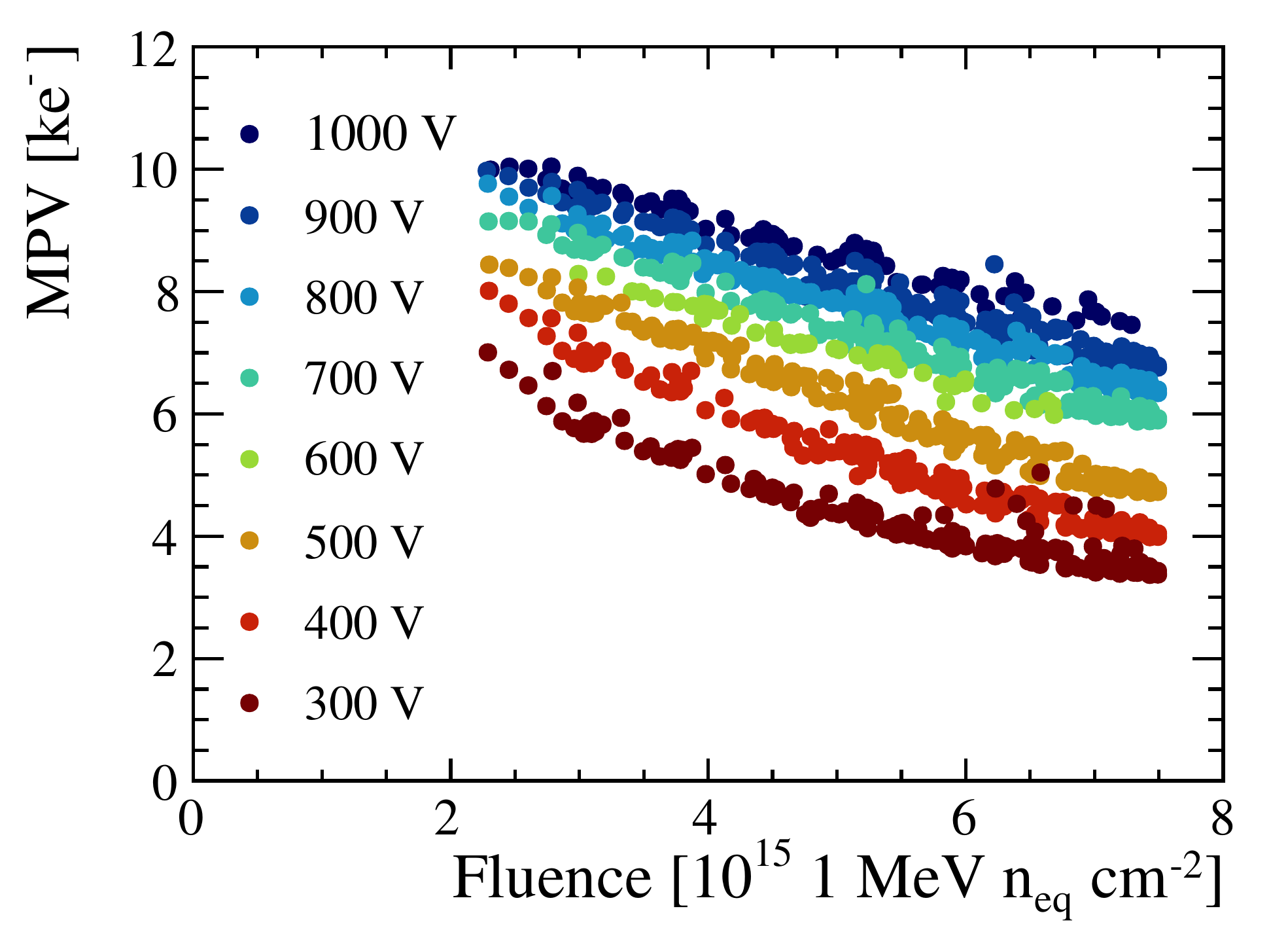}
  \caption{MPV as function of fluence at different bias voltages, 
  after irradiation at IRRAD. Top left: HPK assembly S8 with 35\mum implant width. 
  Top right: HPK assembly S11 with 39\mum implant width. 
  Bottom left: 
  Micron \np assembly S25. Bottom right: Micron \nn assembly S30.} 
  \label{Fig:MPVIRRAD}
\end{figure}

\begin{figure}
  \centering
  \includegraphics[width=0.48\textwidth]{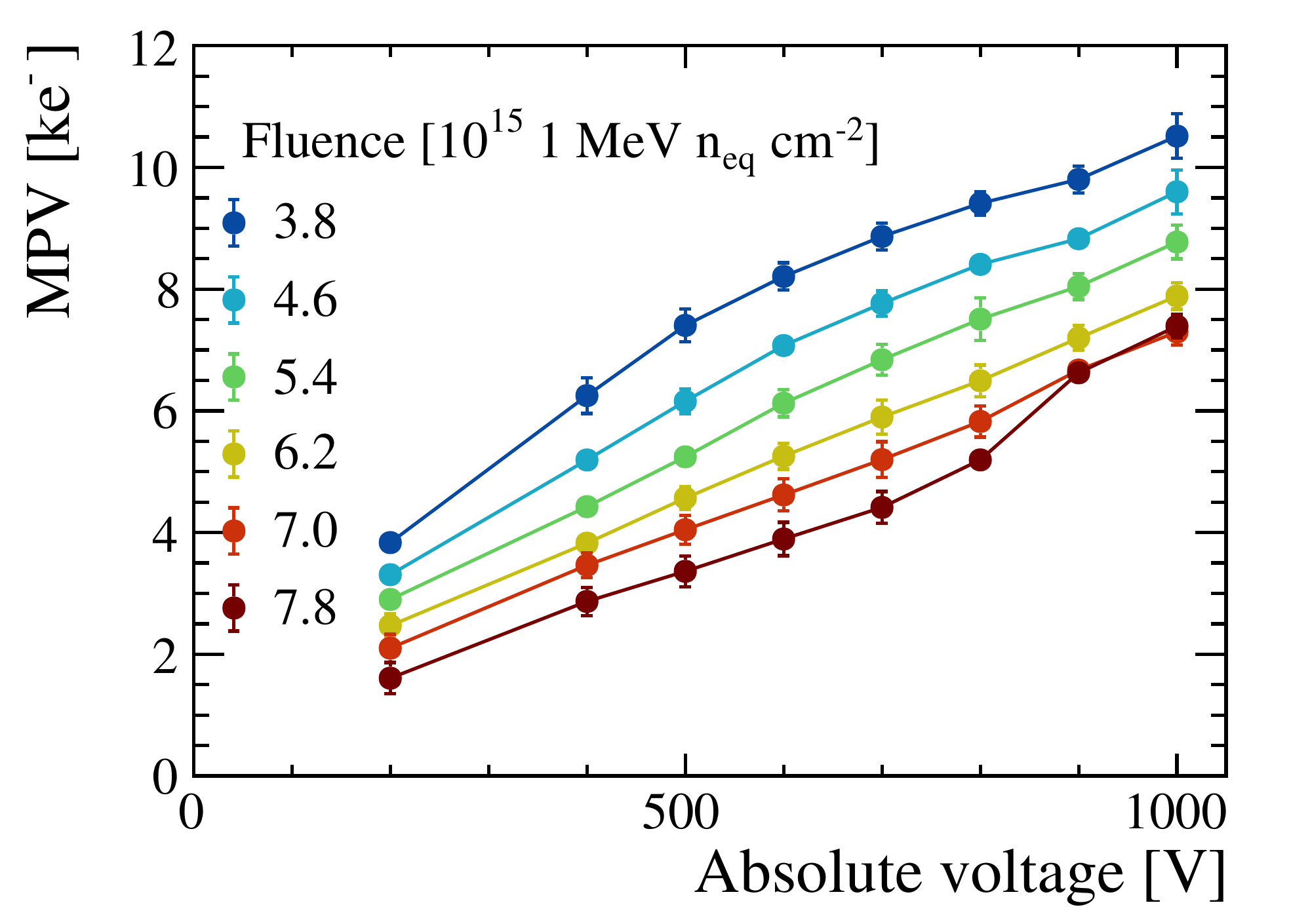}
  \includegraphics[width=0.48\textwidth]{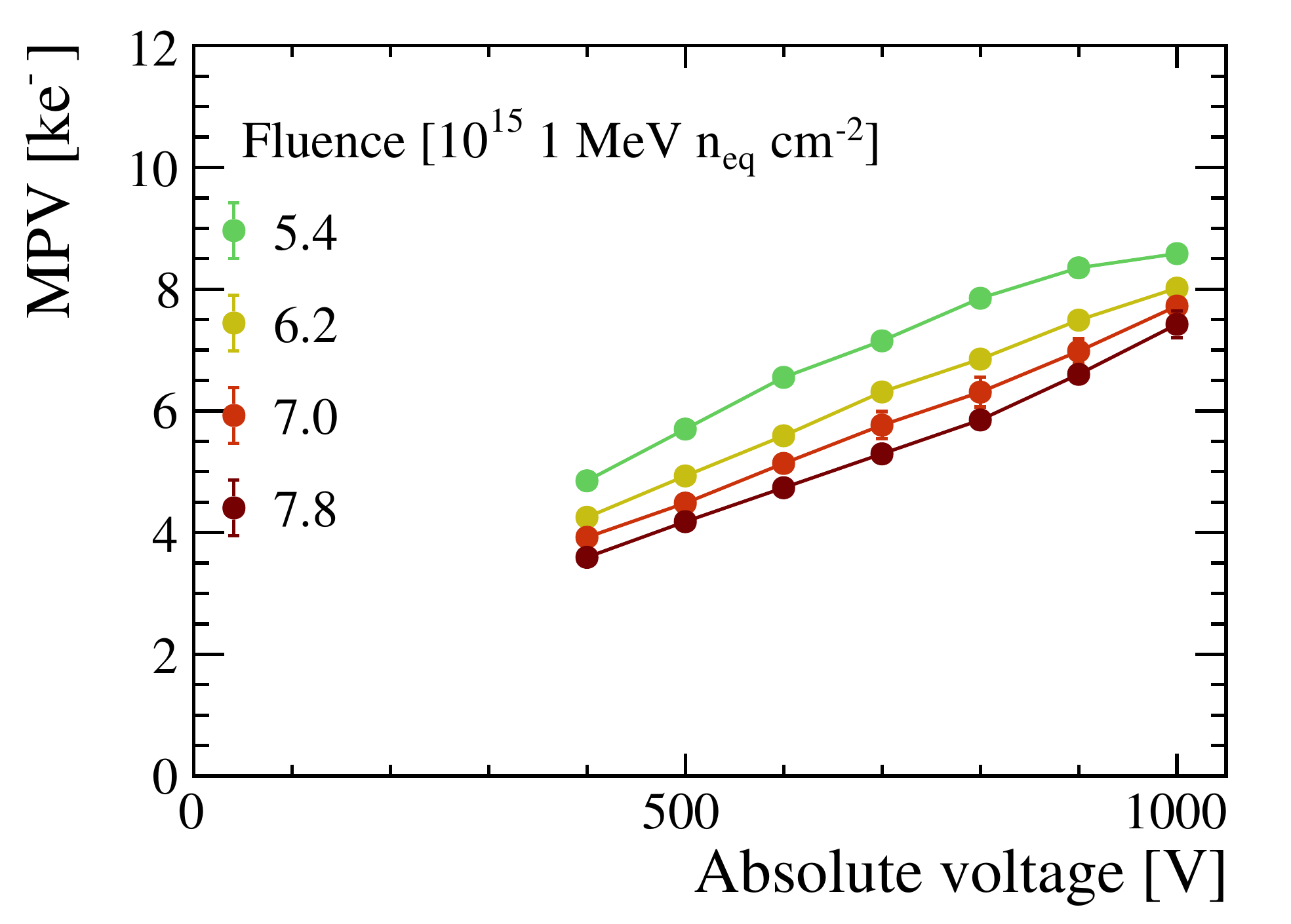}
  \includegraphics[width=0.48\textwidth]{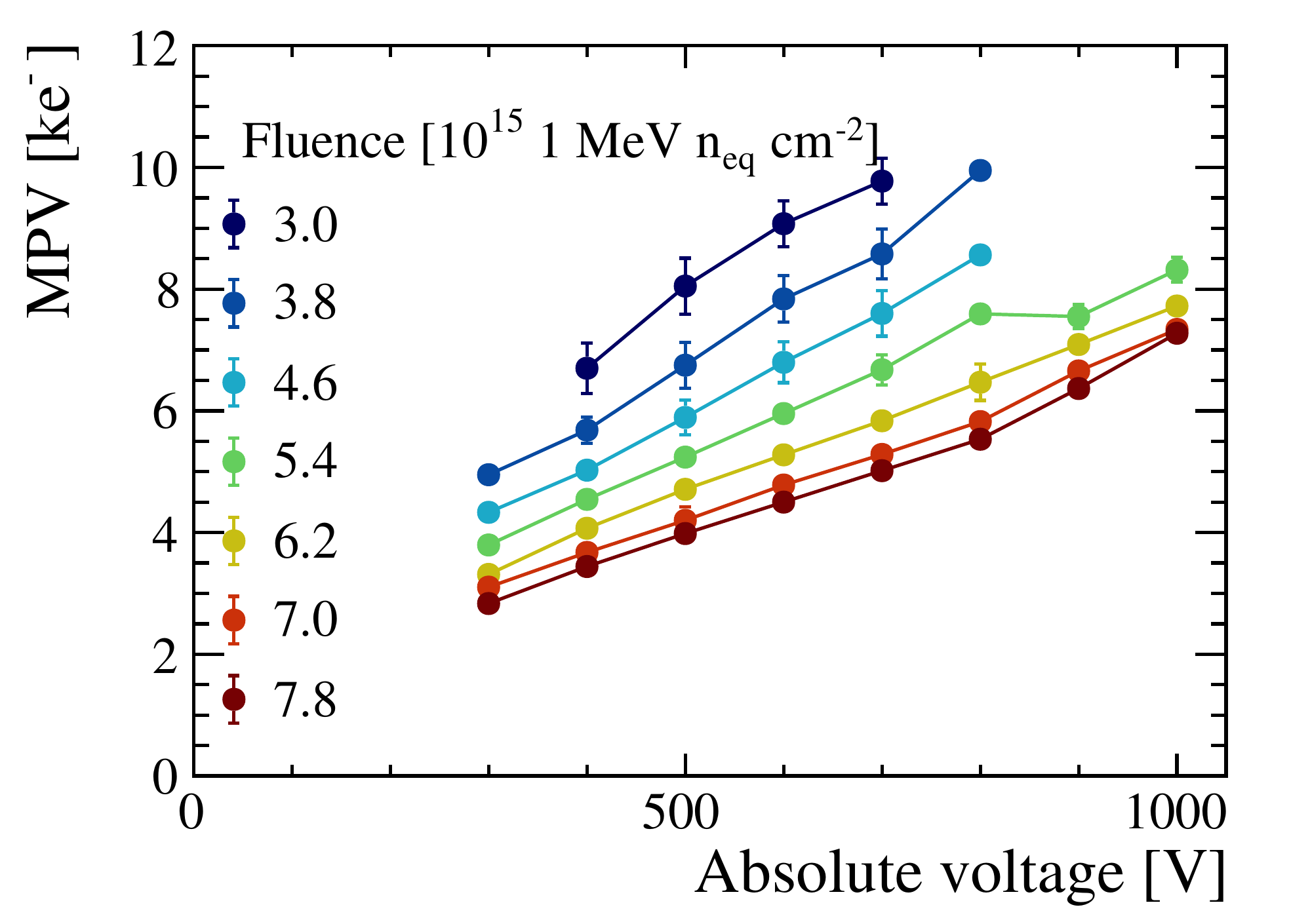}
  \includegraphics[width=0.48\textwidth]{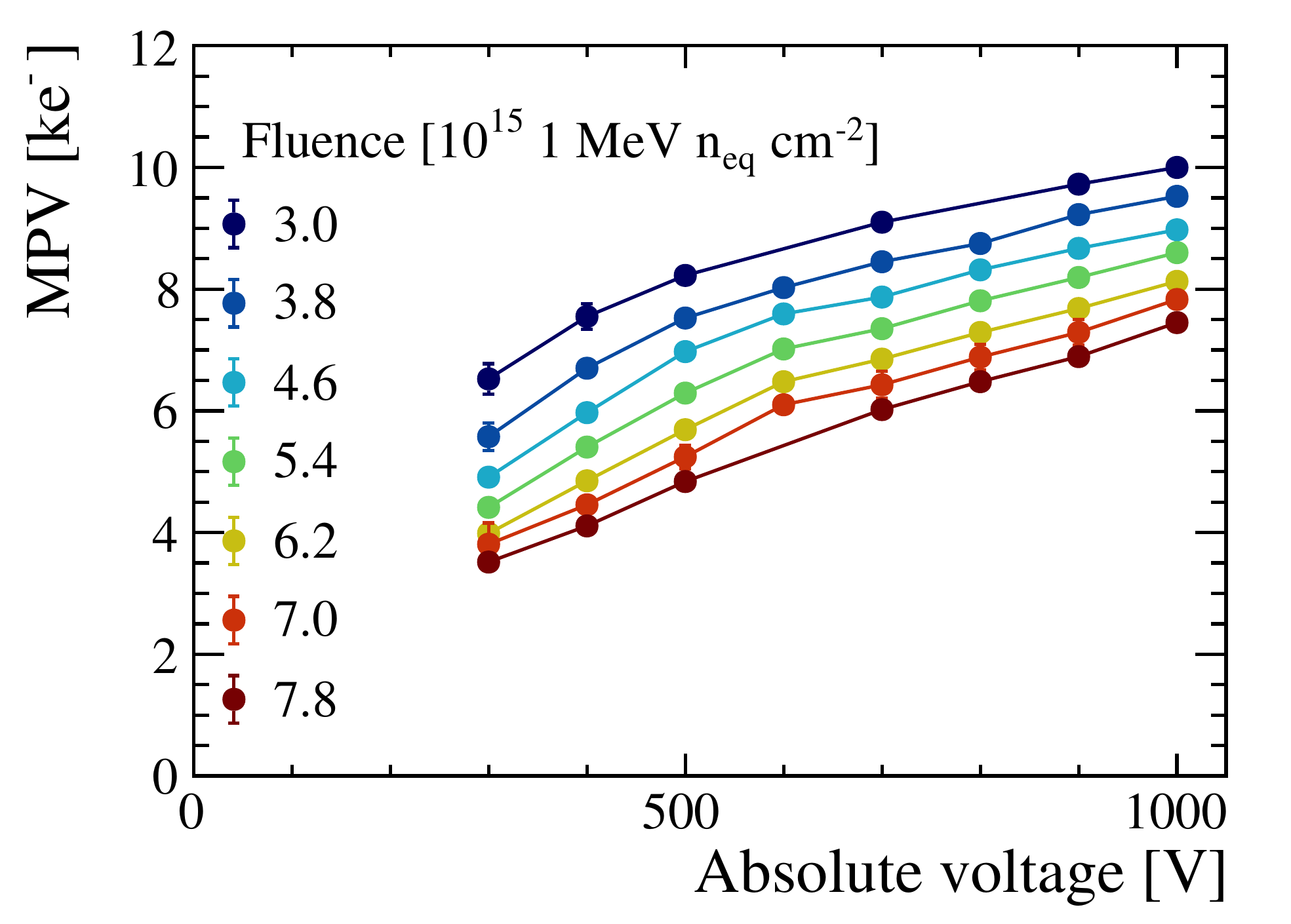}
  \caption{MPV as function of bias voltage at different fluence levels, 
  after irradiation at IRRAD. Top left: HPK assembly S8 with 35\mum implant width. 
  Top right: HPK assembly S11 with 39\mum implant width. 
  Bottom left: 
  Micron \np assembly S25. Bottom right: Micron \nn assembly S30.} 
  \label{Fig:MPVIRRADvoltage}
\end{figure}

\Fig\ref{Fig:MPVIRRAD} shows the MPV as a function of the fluence at different
bias voltages and \fig\ref{Fig:MPVIRRADvoltage} the MPV as a function of the voltage at different fluence levels. Both figures present the results after irradiation at IRRAD.
Charge multiplication is observed in all sensors where the bias voltage can reach \SI{1000}{\volt}, except for the Micron \nn
assembly. This effect emerges at a fluence of about $6.5\times$\fluence and increases
with bias voltage. The difference in the \np and \nn sensors that leads to the presence (absence) in charge multiplication is not known.

\begin{figure}
\centering
\begin{minipage}{.47\textwidth}
\centering
\includegraphics[width=1\linewidth]{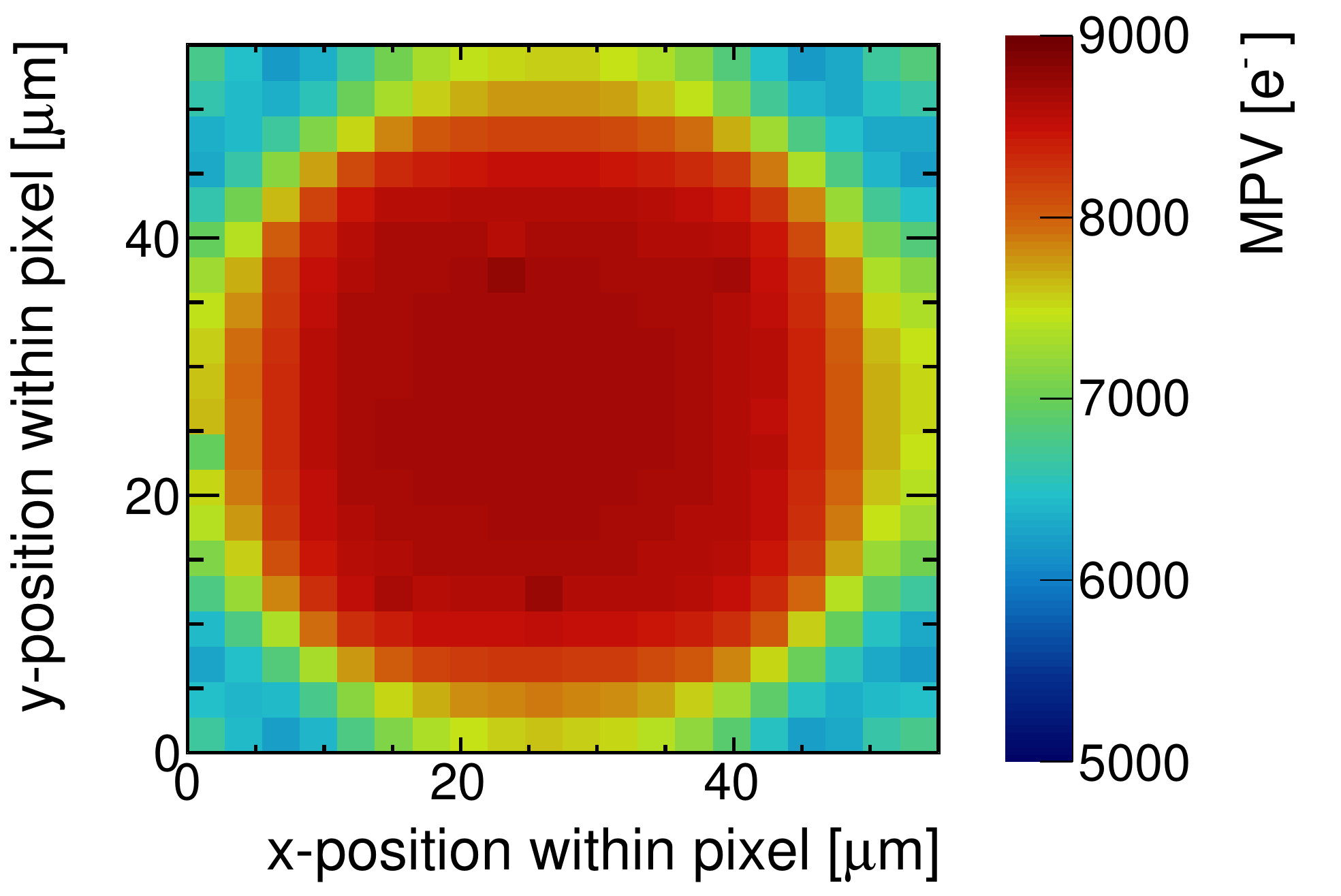}
\end{minipage}\qquad
\begin{minipage}{.47\textwidth}
\centering
\includegraphics[width=1\linewidth]{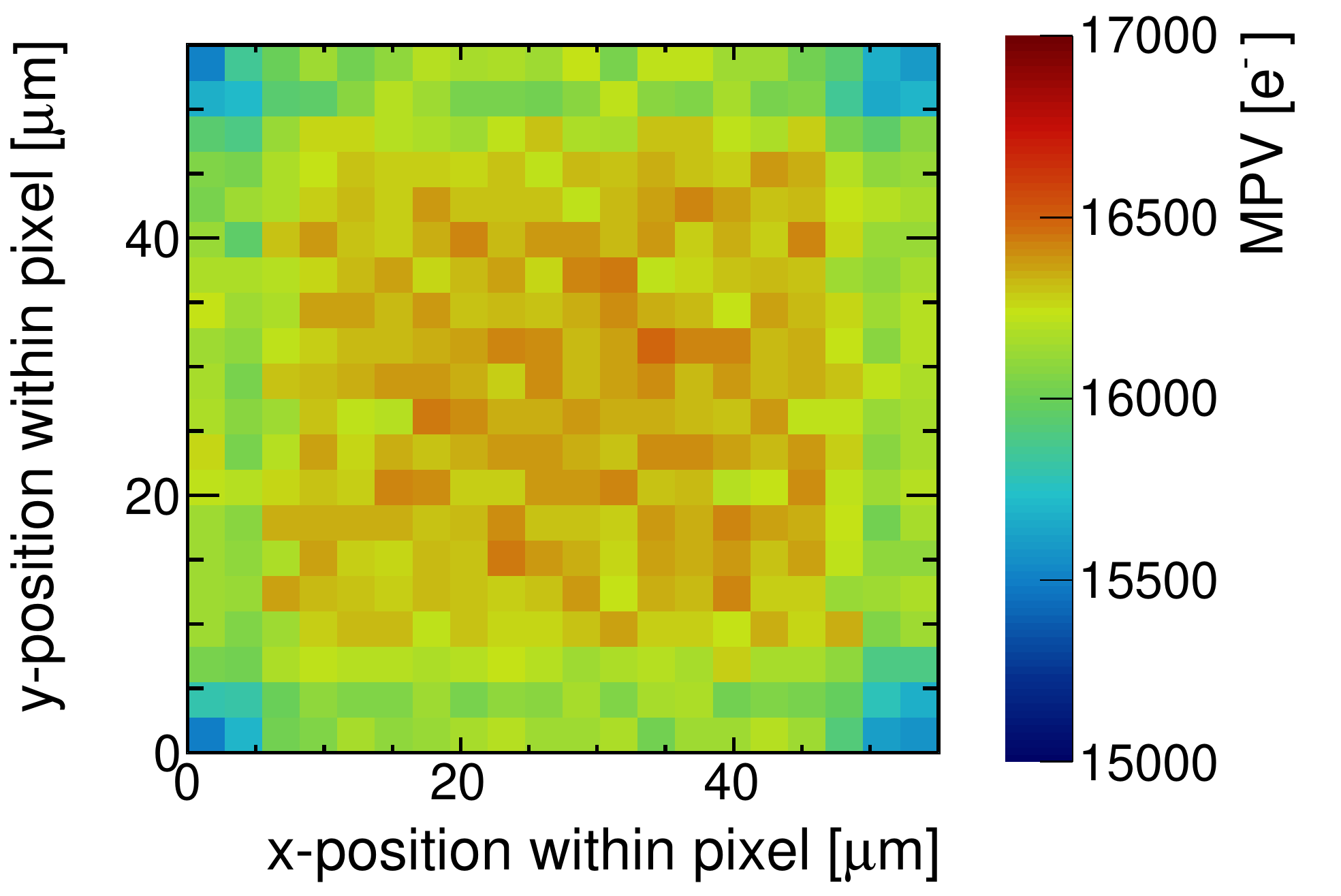}
\end{minipage}

\begin{minipage}[t]{.47\textwidth}
\centering
\captionsetup{width=.99\linewidth}
\captionof{figure}{The MPV of the cluster charge distribution within a pixel for S8 at 1000~V 
after irradiation at IRRAD. Only clusters in the region of the sensor exposed to fluences in the range 7.3 to $7.9\times$\fluence are selected.}
\label{Fig:intraPixel}
\end{minipage}\qquad
\begin{minipage}[t]{.47\textwidth}
\centering
\captionsetup{width=.99\linewidth}
\caption{The MPV of the cluster charge distribution within a pixel for S8 at 800~V before irradiation. }
\label{Fig:intraPixelPre}
\end{minipage}
\end{figure}

Charge multiplication is expected to occur in regions with a high electric field. 
These positions are close to the pixel implant, and thus the amount of charge multiplication could depend on the intrapixel position.
\Fig\ref{Fig:intraPixel} shows the MPV for different intrapixel positions 
of assembly S8 after irradiation at IRRAD up to maximum fluence at a bias of \SI{1000}{\volt}. 
For this figure, clusters are selected in the regions of the sensor that were exposed to fluences in the range of 7.3 to $7.9\times$\fluence. 
As a comparison, the intrapixel MPV for S8 before irradiation is shown in 
\fig\ref{Fig:intraPixelPre}. 
Diagonal bands can be seen near the corners of the pixel; here the MPV is lower than in the corner itself. This is due to charge sharing 
with neighbouring pixels. 
A detailed explanation is given in \app\ref{sec:appendixClCharge}.

\begin{figure}
\centering
\begin{minipage}{.47\textwidth}
\centering
\includegraphics[width=1\linewidth]{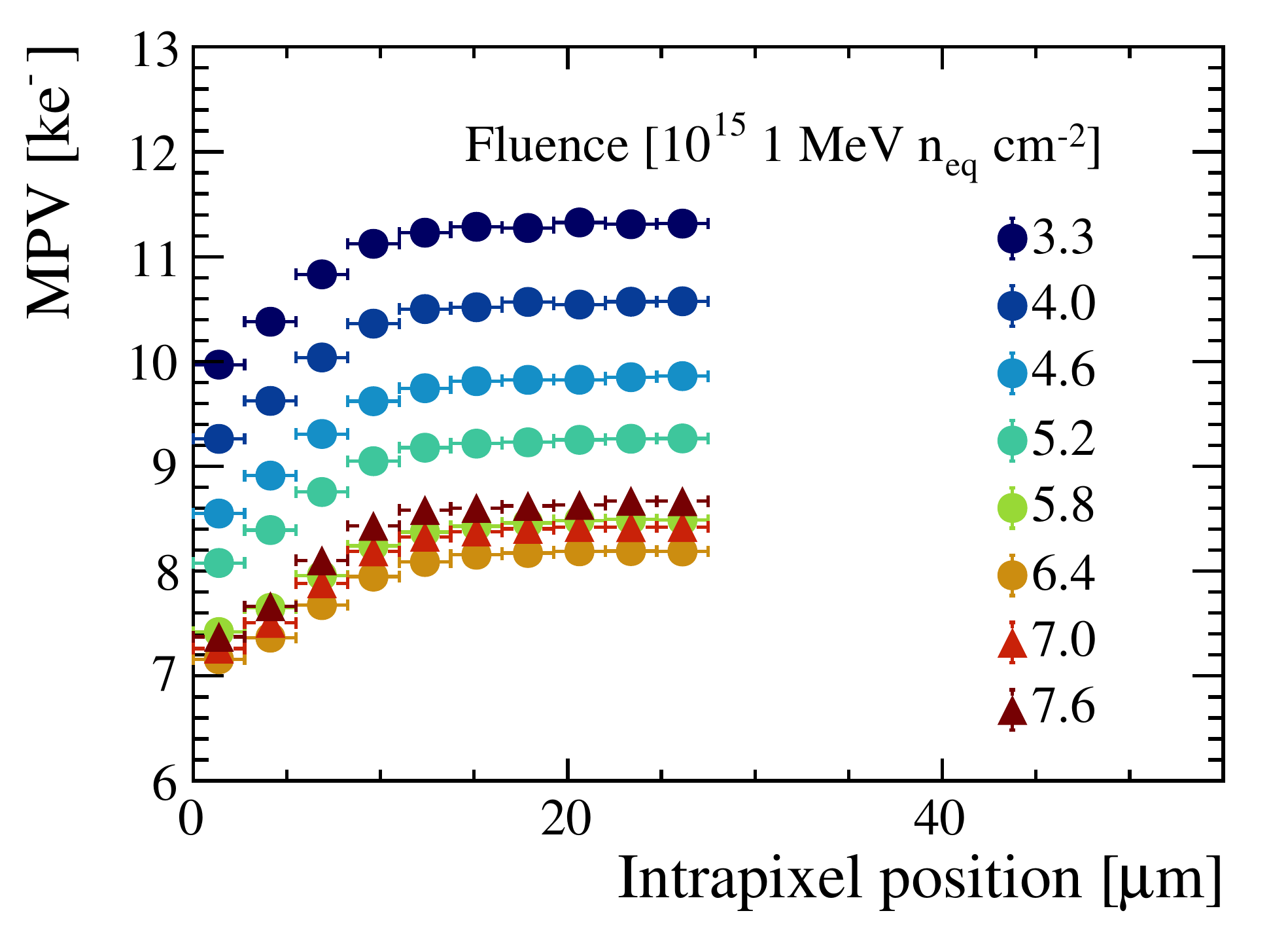}
\end{minipage}\qquad
\begin{minipage}{.47\textwidth}
\centering
\includegraphics[width=1\linewidth]{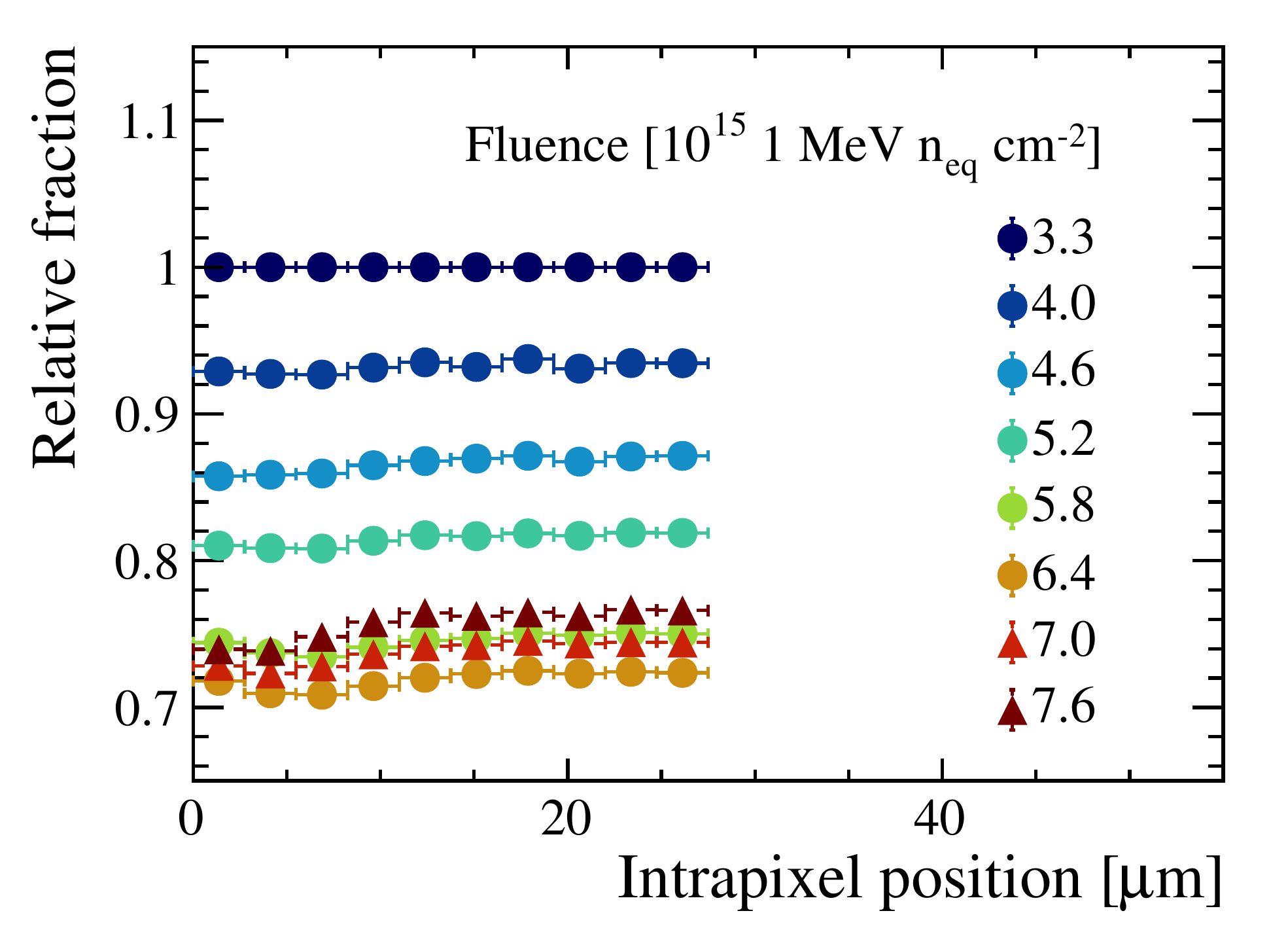}
\end{minipage}


\begin{minipage}[t]{.47\textwidth}
\centering
\captionsetup{width=.99\linewidth}
\captionof{figure}{Intrapixel cluster charge MPV  (for $y=27.5$\mum) for different 
fluence values of S8 at an operating voltage of 1000~V. }
\label{Fig:intraPixelCharge}
\end{minipage}\qquad
\begin{minipage}[t]{.47\textwidth}
\centering
\captionsetup{width=.99 \linewidth}
\caption{The relative fraction of the intrapixel charge in comparison to the lowest fluence for the cross-section of the pixel at $y=27.5$\mum. }
\label{Fig:intraPixelRelative}
\end{minipage}
\end{figure}

For the IRRAD assemblies, the intrapixel MPV study is performed for different fluence values up to maximum fluence. 
By comparing a slice of $5.5\mum$ along the middle of the pixel ($y=27.5 \mum$) of the intrapixel cluster charge distribution at different fluence levels, the position within the pixel where charge multiplication occurs can be determined, as is shown in  
\Fig\ref{Fig:intraPixelCharge}. The MPV decreases with fluence up to $7\times\fluence$. 
At this point the MPV begins to increase with fluence, which is a clear indication that charge multiplication is occurring. 
The small drop from the middle to the edge of the pixel is due to charge sharing effects.
The charge collected relative to that of the lowest fluence bin is shown in \fig\ref{Fig:intraPixelRelative} in order to demonstrate where in the pixel the charge multiplication occurs.
It can be concluded that charge multiplication occurs uniformly
over the centre of the pixel, while towards the edge of the pixels charge multiplication seems to be lower.

\subsection{Edge performance}
\label{sec:edge}

The performance of the sensors at the edge, near the guard ring region, is investigated by illuminating these areas with the beam.
Results are shown for one corner of the sensor, but it has been verified that the other corners behave the same.
Nonirradiated sensors are operated at a bias where they are above the depletion voltage estimated from charge saturation measurements.
The results are shown for single chip sensors, but all prototypes tested exhibit consistent behaviour at the edge between single and triple sensors as well as between different types of irradiation. 

\begin{figure}[h!]
  \centering
  \includegraphics[width=0.75\textwidth]{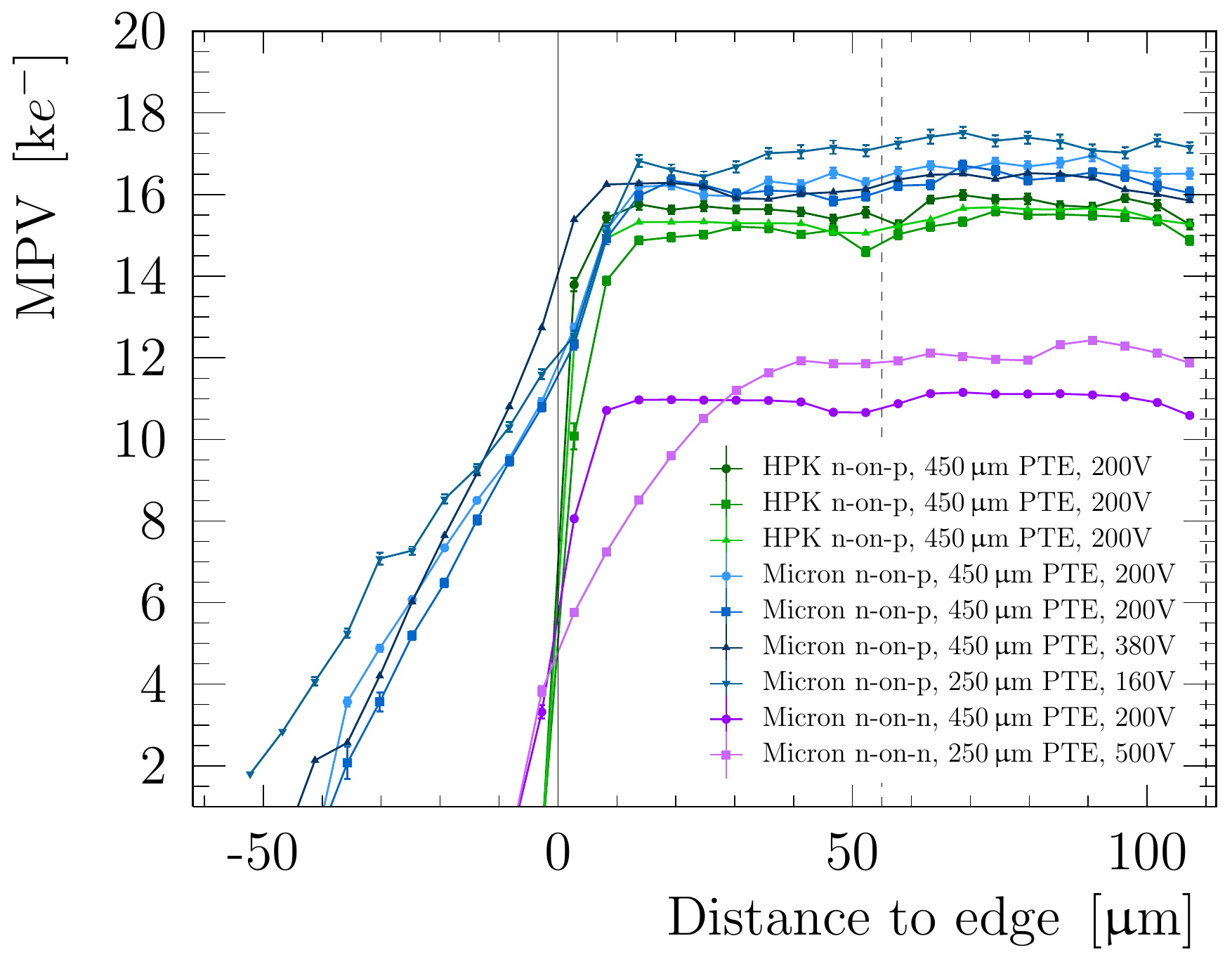}
  \caption{Cluster charge MPV as a function of the distance of the track intercept from the edge of the pixel matrix for all the nonirradiated assemblies tested. The solid line represents the border of the pixel matrix and the dashed lines the borders between pixels. }
  \label{fig:summary_non_irr}
\end{figure}

The performance at the edge of the pixel matrix is quantified by measuring the cluster charge as a function of the track intercept position.
The expected behaviour is to collect charge from tracks traversing the pixel matrix, but not from tracks passing through the guard ring area apart from a small amount of diffused charge.
The MPV as a function of the track distance from the edge in the $x$ direction (corresponding to increasing column numbers) can be seen in \fig\ref{fig:summary_non_irr} for all the tested nonirradiated assemblies.
The uncertainty assigned to the MPV is statistical only.
The solid line represents the edge of the pixel matrix and the dashed lines the borders between pixels.
Two prototype variants exhibit an undesired behavior. All of the $200\mum$ Micron \np sensors accumulate charge in the last pixel from tracks traversing the sensor beyond the edge of the pixel matrix, while the $150\mum$ Micron \nn (250\mum PTE backside guard rings) sensor does not collect the full charge in the last pixel.
None of the HPK sensors show this behaviour apart from some expected charge diffusion at the boundary between the guard ring and the edge pixel.

\begin{figure}[h!]
    \centerline{\includegraphics[width=0.85\textwidth]{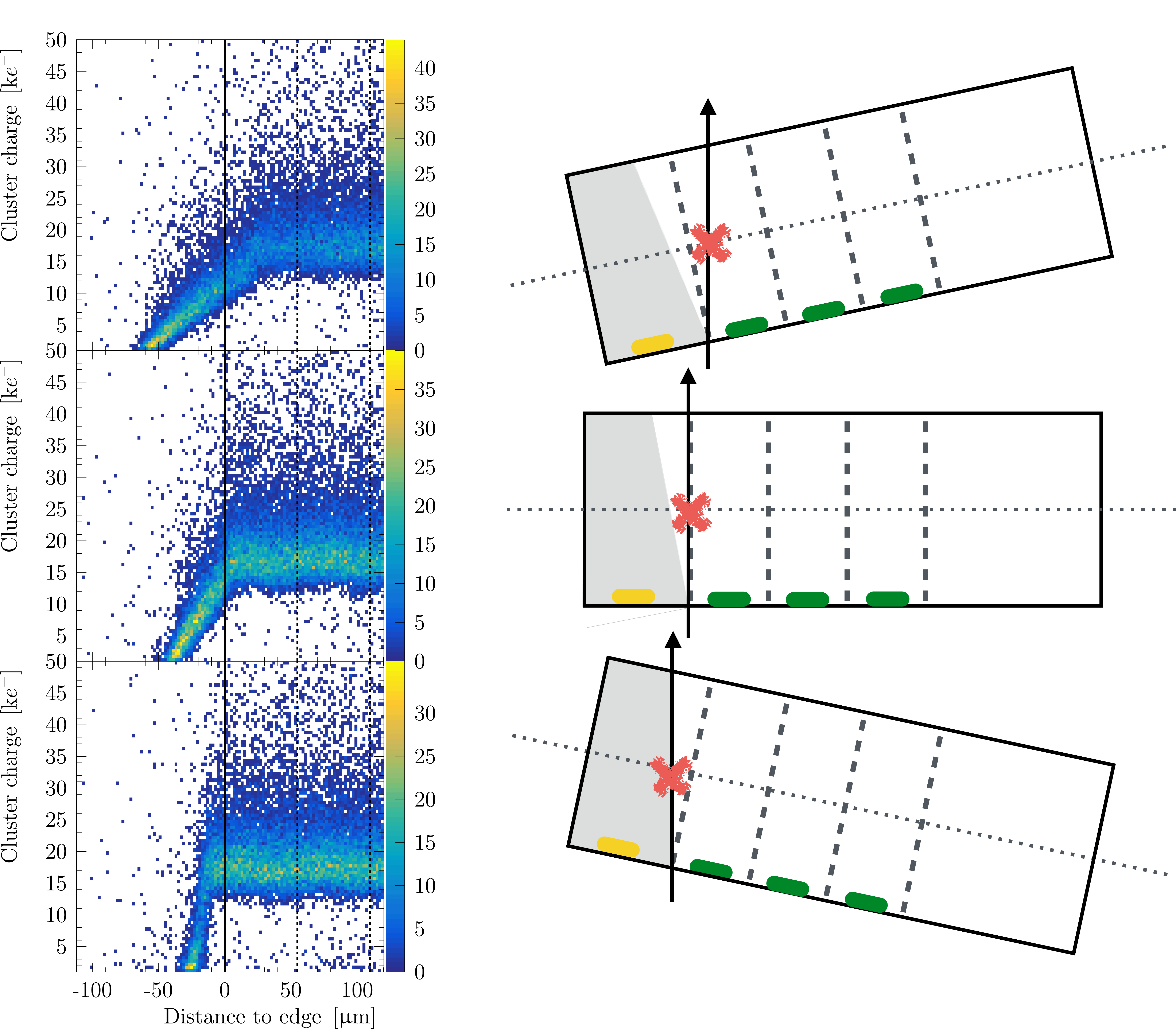}}
    \caption{Cluster charge as a function of the distance of the track intercept from the edge (left) and related sketch (right) for a $200 \mum$ thick Micron \np sensor with $450 \mum$ guard rings operated at 380~V. The sensor is rotated with respect to the beam by $-12\degrees$ (top), $0\degrees$ (middle) and $+12\degrees$ (bottom).}
    \label{fig:sketchnp}
\end{figure}

The simplest geometrical model that can explain the observed edge effect for Micron \np and Micron \nn sensors is a tilted border between the charge collection region of the guard rings and that of the pixels.
This is illustrated in \fig\ref{fig:sketchnp} for  Micron \np sensors, where the charge distribution as a function of the  track intercept position, $x$, is studied at three different angles: -12$^{\circ}$, 0$^{\circ}$ and +12$^{\circ}$.
In the schemes, the black arrow represents the particle trajectory nearest to the physical edge where the deposited charge is fully collected, with the red cross showing the corresponding measured position of the intercept between the track and the sensor, and the shaded area indicating the charge collection region of the guard rings.
At -12$^{\circ}$ the slope of the charge deposit is present and extends far beyond the edge of the matrix. 
The charge is fully collected by the pixel implants up to the point indicated  by the red cross. 
Beyond that point the first pixel starts to pick up charge from the guard ring region (represented in grey).
At +12$^{\circ}$ the slope in the charge deposit almost disappears. The charge is either fully collected by the pixel implant or fully collected by the guard rings.
This model explains why in the case of Micron \np sensors charge is collected from the guard ring region at the pixel side.
Analogously, angled measurements with Micron \nn sensors show that the border is tilted in the opposite direction, such that charge from the outermost pixel is lost to the guard rings at the backplane side~\cite{DallOccoThesis}.
The tilted border hypothesis could be verified by a TCAD\footnote{Technology Computer Aided Design} simulation, but this was not pursued because the details of the guard ring design is proprietary information.

For Micron \np sensors, the observed edge effect leads to an approximately $30\%$ excess of hits in the first column of the pixel matrix 
with the sensor biased above depletion voltage
and traversing tracks at normal incidence.
The second column of the pixel matrix also shows an excess of hits, leading to the conclusion that the distortion of the electric field is such that the second column collects charge from tracks going through the first column.
A similar effect has been observed for active-edge silicon sensors in ref.~\cite{Akiba:2014mna}.
The excess of hits decreases with increasing operation voltage due to the stronger electric field, and it is also observed to decrease when tracks are at an angle with respect to the sensor.
An excess of hits in the first column and row of the pixel matrix is also observed in a lab experiment using a $^{90}$Sr source, verifying the beam test findings.
Data have been collected operating the sensor at different voltages and the observed hit excess dependence has the same trend as the one obtained from testbeam data.

\begin{figure}[h!]
  \centering
  {\includegraphics[width=0.75\textwidth]{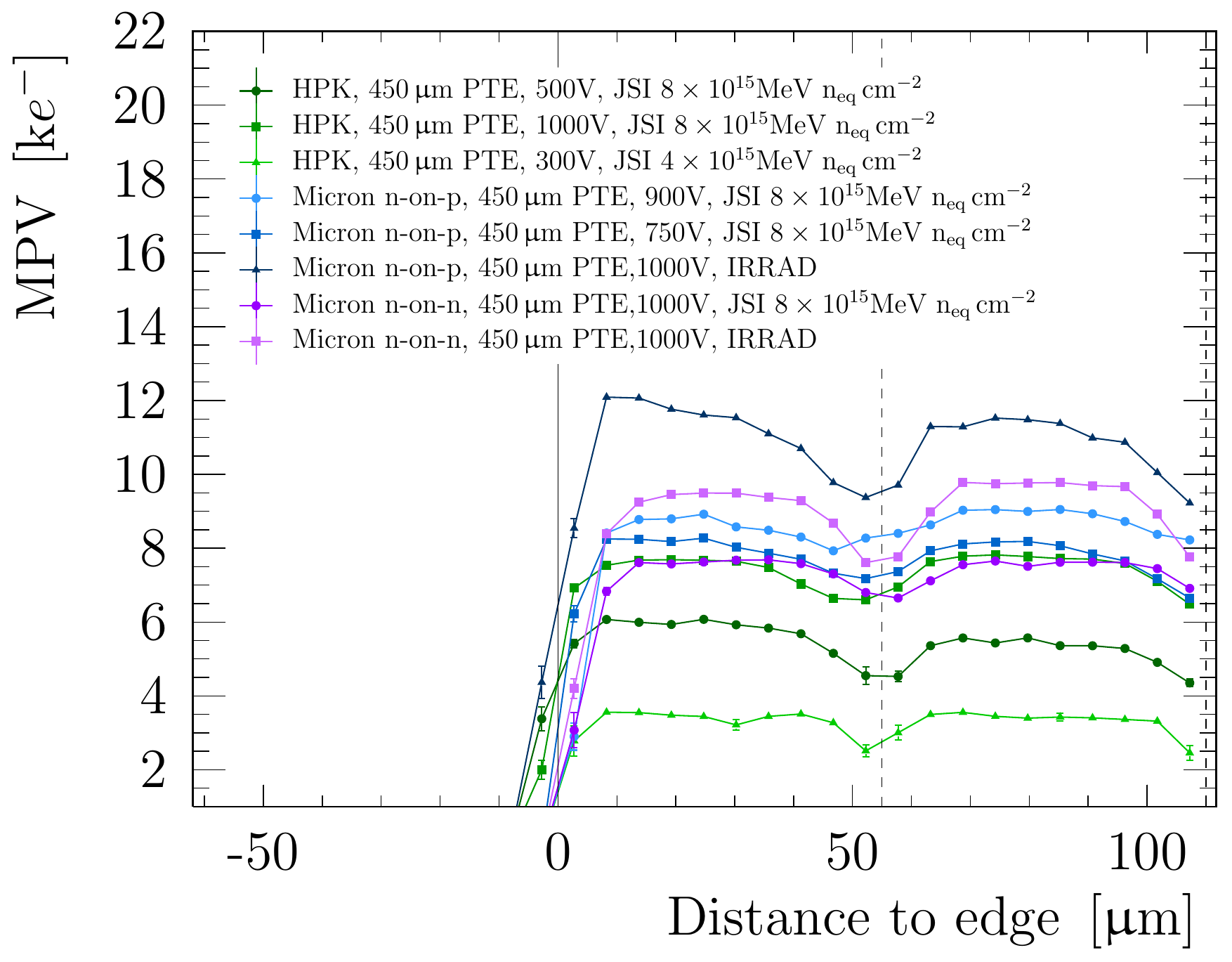}}
  \caption{MPV of the cluster charge as a function of the distance to the edge of the associated track intercept for all the assemblies uniformly irradiated at JSI and nonuniformly    irradiated at IRRAD. The solid line represents the border of the pixel matrix and the dashed lines the borders between pixels.}
\label{fig:summary_irr_all}
\end{figure}

The MPV profiles for irradiated sensors are shown in \fig\ref{fig:summary_irr_all} as a function of the distance to the edge in the $x$ direction.
After irradiation by either protons or neutrons with fluences ranging from 2 to $8\times$\fluence, the behaviour at the edge is comparable among the different prototype variants.
The linear charge deposit extending beyond the edge observed in nonirradiated Micron \np sensors is strongly reduced after irradiation.
This can be explained by the fact that the sensors are not fully depleted even at the  highest operational voltage used in the tests.

\section{Conclusions and outlook}
\label{sec:conclusion}

An extensive testing campaign has been performed for the VELO upgrade, investigating several planar silicon sensor designs from two different vendors: HPK and Micron.
In this paper, a study of the charge collection before and after irradiation up to \maxfluence of these sensors is presented. Irradiation of the sensors was performed at JSI with neutrons and at IRRAD and KIT with protons. 
The MPV as a function of bias voltage is determined before irradiation, and shows saturation at 16\unit{ke^-}, 17\unit{ke^-}, and 12\unit{ke^-} for HPK \np, Micron \np, and Micron \nn, respectively. 
After irradiation, the MPV decreases significantly and shows a linear dependence on the bias voltage. 
At a bias voltage of \SI{1000}{\volt}, the MPV is around 8\unit{ke^-}, indicating that the sensor achieves the requirement of collecting more than 6\unit{ke^-} after irradiation to the maximum expected fluence for the VELO upgrade.

During the campaign charge multiplication was observed for both the HPK and Micron \np sensors, while no charge multiplication is observed for the Micron \nn sensors.
Charge multiplication is observed uniformly over the centre of the pixel, while it seems less near the edge.

Two classes of tested prototypes exhibited unexpected behaviour at the edge prior to irradiation.
In the case of Micron \np sensors, charge is gained from the guard ring region at the pixel implant side, while in the case of the Micron \nn sensors with $250 \mum$ PTE, charge from the outermost pixels is lost to the guard ring area.
In view of the VELO upgrade the observed edge effect is critical, since it would increase the occupancy in the part of the sensor where it is already the highest.
Moreover, it leads to a loss of spatial resolution for the first measured point.

The studies presented in this paper were decisive in the choice to adopt the 200\mum thick HPK \np sensors with 39\mum implant width and 450\mum guard ring size for the VELO upgrade.

\section*{Acknowledgments}

We would like to express our gratitude to our colleagues in the CERN accelerator departments for the excellent performance of the beam in the SPS North Area. 
We would like to thank  
Angelo Di Canto,
Stefano De Capua,
Xabier Cid Vidal,
Álvaro Dosil Suárez,
Antonio Fernández Prieto,
Julian Garcia Pardinas,
Dawid Gerstel,
Christoph Hombach,
Eddy Jans,
Suzanne Klaver,
Carlos Vázquez Sierra,
Maria Vieites Diaz,
and Heather Wark 
for taking part in the data taking effort throughout the years of 2014 to 2016.
We would also like to express our gratitude towards the institutes that irradiated the sensors used in this paper: IRRAD, JSI, and KIT.
We gratefully acknowledge the financial support from CERN and from the national agencies: CAPES, CNPq, FAPERJ (Brazil); the Netherlands Organisation for Scientific Research (NWO); The Royal Society and the Science and Technology Facilities Council (U.K.); MNiSW and NCN UMO-2018/31/B/ST2/03998 (Poland). This project has received funding from the European Union’s Horizon 2020 Research and Innovation programme under Grant Agreement no. 654168.

\clearpage

\appendix

\section{List of assemblies}
\label{sec:appendixAssemblies}

The details of the assemblies that are tested during this analysis are summarised here. Two different types of substrates (\np and \nn) are used from two different vendors. Further differences between individual assemblies are indicated in the table below. 

\begin{table}[h]
\footnotesize
\centering
\caption{Assemblies tested during the charge collection property studies. Single-chip assemblies are indicated by an S in their ID, while triple-chip assemblies are indicated by a T.}
  \begin{tabular}{lccccccc}
  \hline
  ID   & Vendor & \makecell{Thickness \\ {[$\!\mum$]} } & Type  & \makecell{Edge width \\  {[$\!\mum$]} } & \makecell{Implant \\  {[$\!\mum$]} } & \makecell{Irradiation \\ facility } & \makecell{Peak fluence \\  {[\fluence]} } \\
  \hline 

  S4  & HPK   & 200 & \np & 600 & 39 & KIT & 4 \\
  S6  & HPK   & 200 & \np & 450 & 39 & JSI & 8 \\
  S8  & HPK   & 200 & \np & 450 & 35 & IRRAD & 8 \\
  S11  & HPK   & 200 & \np & 450 & 39 & IRRAD & 8 \\
  S15  & HPK   & 200 & \np & 450 & 35 & JSI & 4 \\
  S17  & HPK   & 200 & \np & 450 & 39 & JSI & 8 \\
  S18  & HPK   & 200 & \np & 450 & 39 & - & - \\
  S20  & HPK   & 200 & \np & 450 & 35 & - & - \\
  S22  & HPK   & 200 & \np & 450 & 35 & JSI & 8 \\
  S23  & Micron   & 200 & \np & 450 & 36 & JSI & 8 \\
  S24  & Micron   & 200 & \np & 450 & 36 & JSI & 8 \\
  S25  & Micron   & 200 & \np & 450 & 36 & IRRAD & 8 \\
  S27  & Micron   & 150 & \nn & 450 & 36 & JSI & 8 \\
  S29  & Micron   & 150 & \nn & 450 & 36 & JSI & 8 \\
  S30  & Micron   & 150 & \nn & 450 & 36 & IRRAD & 8 \\
  S31  & Micron   & 200 & \np & 250 & 36 & - & - \\
  S33  & Micron   & 150 & \nn & 250 & 36 & - & - \\
  \hline
  T2  & HPK   & 200 & \np & 450 & 35 & KIT & 4 \\
  T15  & Micron   & 200 & \np & 450 & 36 & KIT & 8 \\
  T23  & HPK   & 200 & \np & 450 & 39 & - & - \\

  \hline
  \end{tabular}
    \label{tab:assemblies}
\end{table}

\section{Shape of the intrapixel cluster charge distribution}
\label{sec:appendixClCharge}

\begin{figure}[h]
  \centering
  {\includegraphics[width=0.47\textwidth]{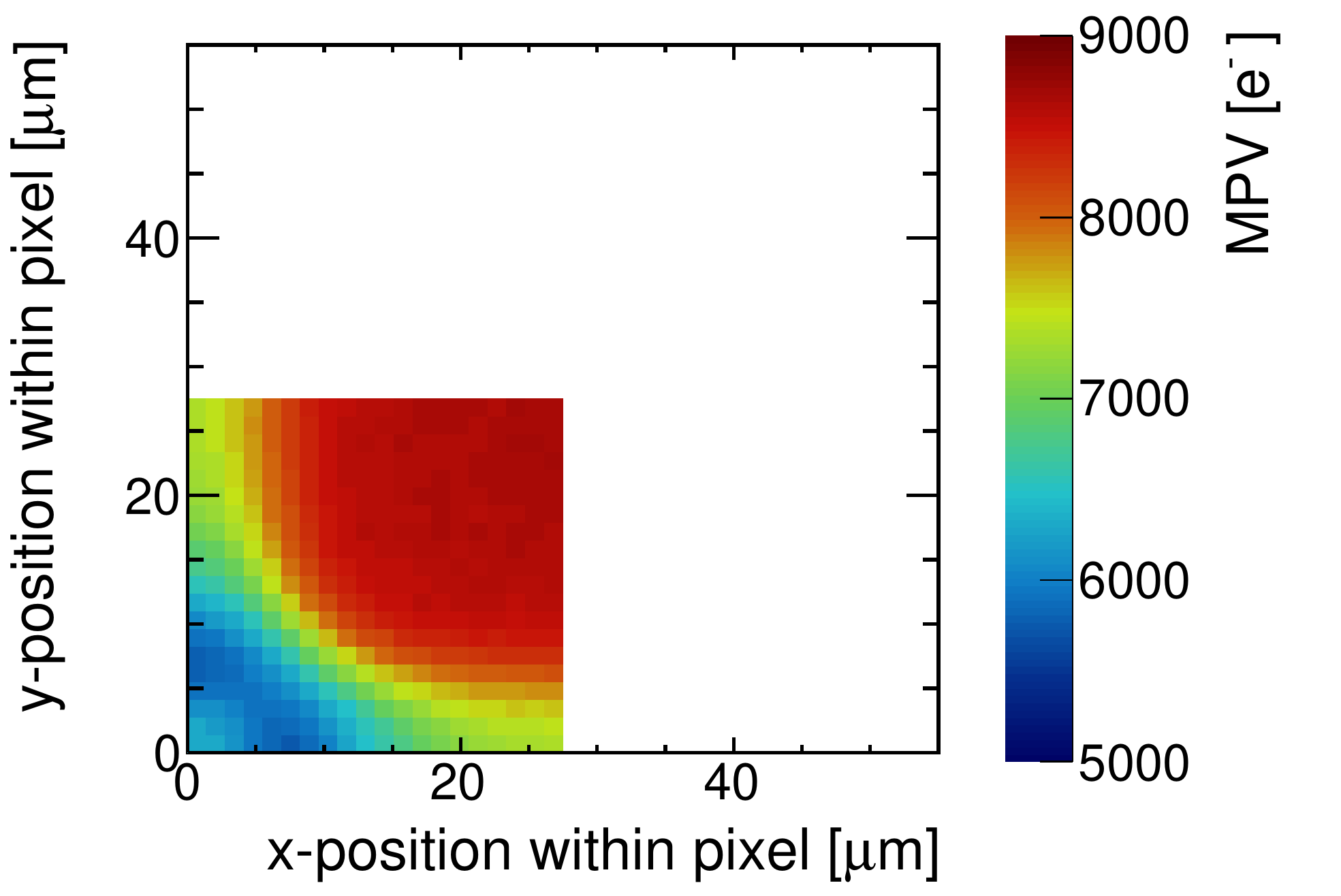}}  
  \caption{The MPV of the cluster charge distribution within a pixel for S8 at 1000~V 
after irradiation at IRRAD. Only clusters in the fluence range of 7.3 to $7.9\times$\fluence are selected.}
  \label{fig:AppendixAllClusterSize}
\end{figure}

The MPV distribution within a single pixel shows an peculiar feature: a diagonal region near the corner is observed to have a lower MPV than both the centre and the corners of the pixel, as is shown in \fig\ref{fig:AppendixAllClusterSize}. For this figure, the four corners of pixel have been `folded' into one in order to maximise the effective size of data sample.

\begin{figure}[!p]
  \centering
  {\includegraphics[width=0.47\textwidth]{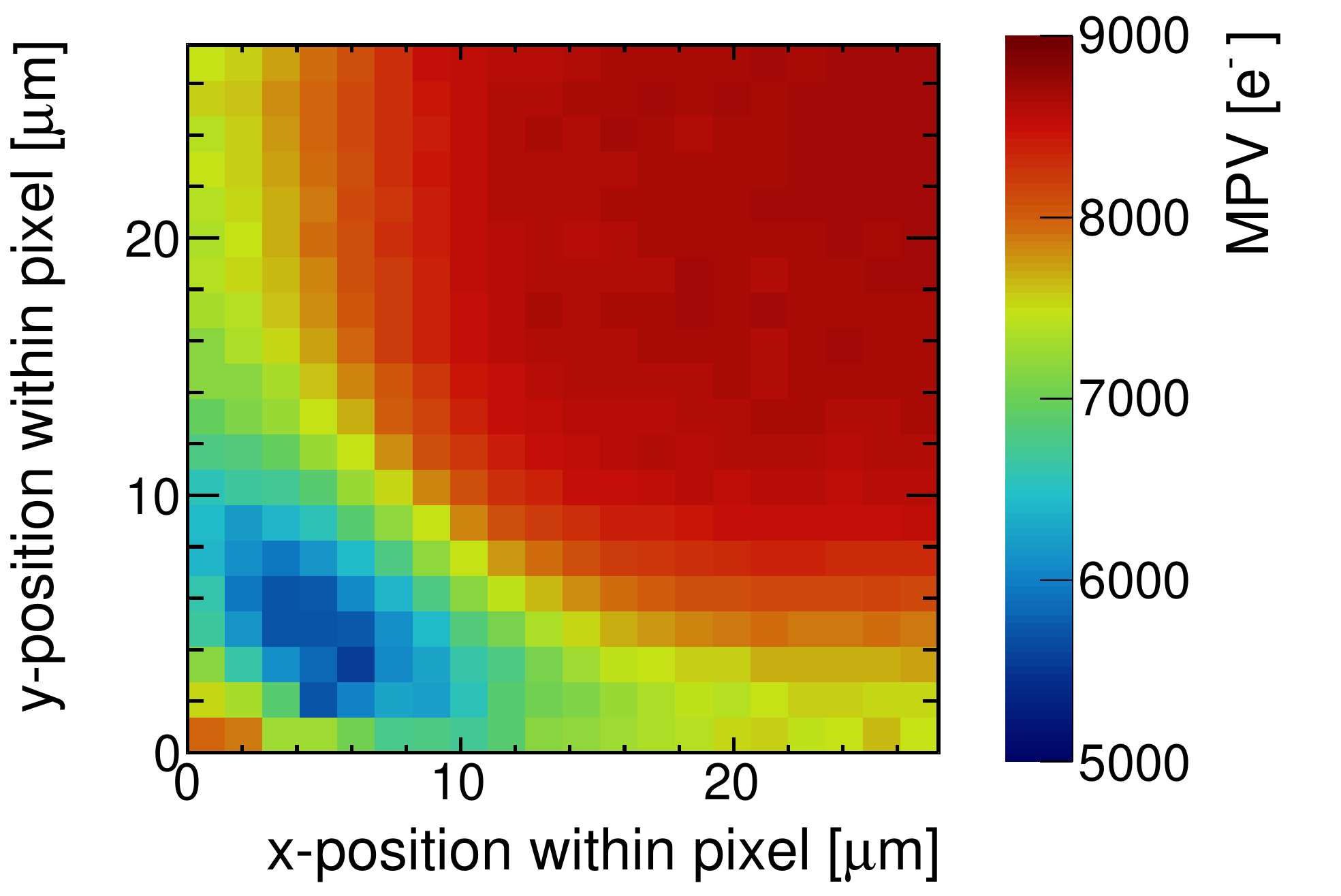}} \quad
  {\includegraphics[width=0.47\textwidth]{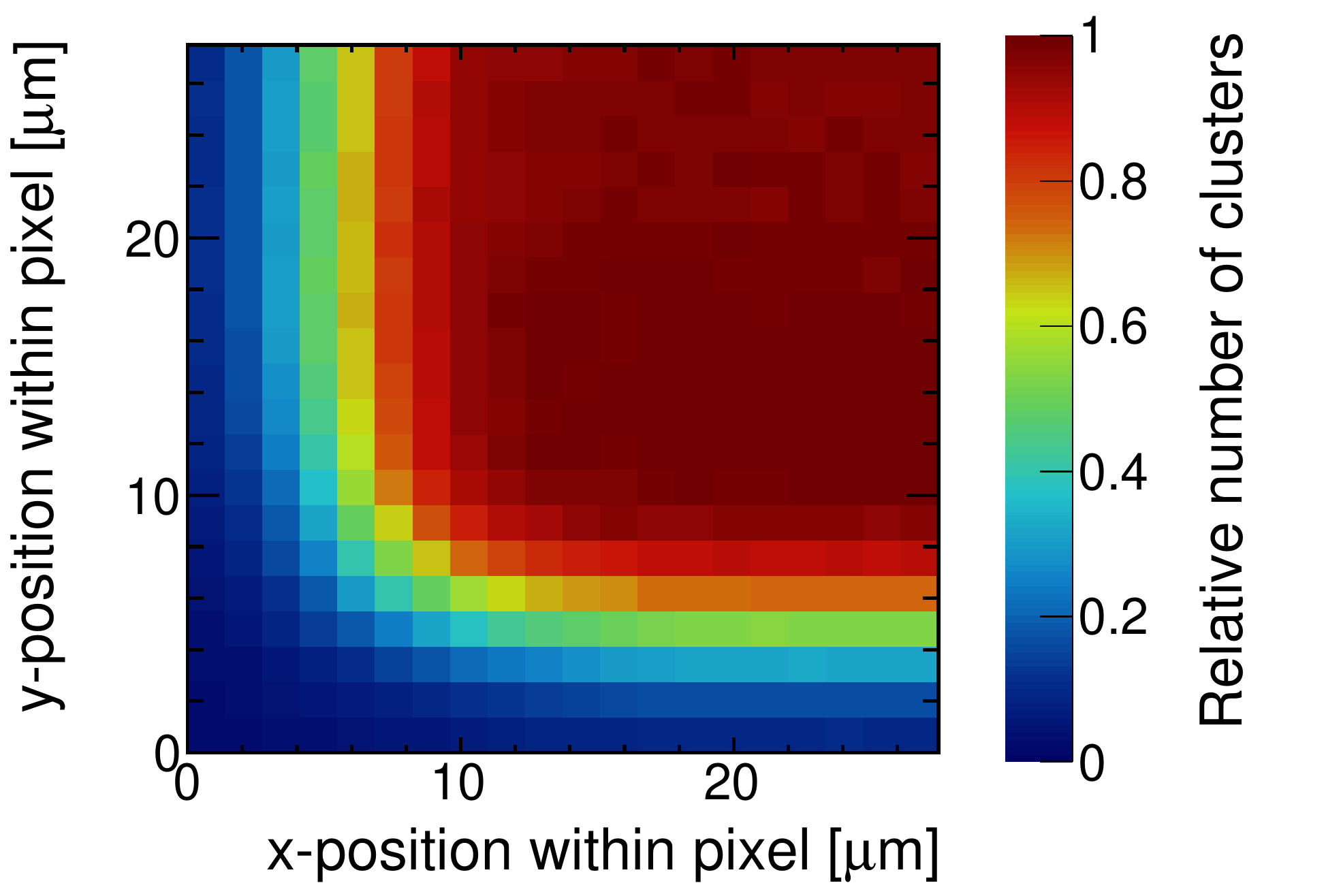}} \\
  {\includegraphics[width=0.47\textwidth]{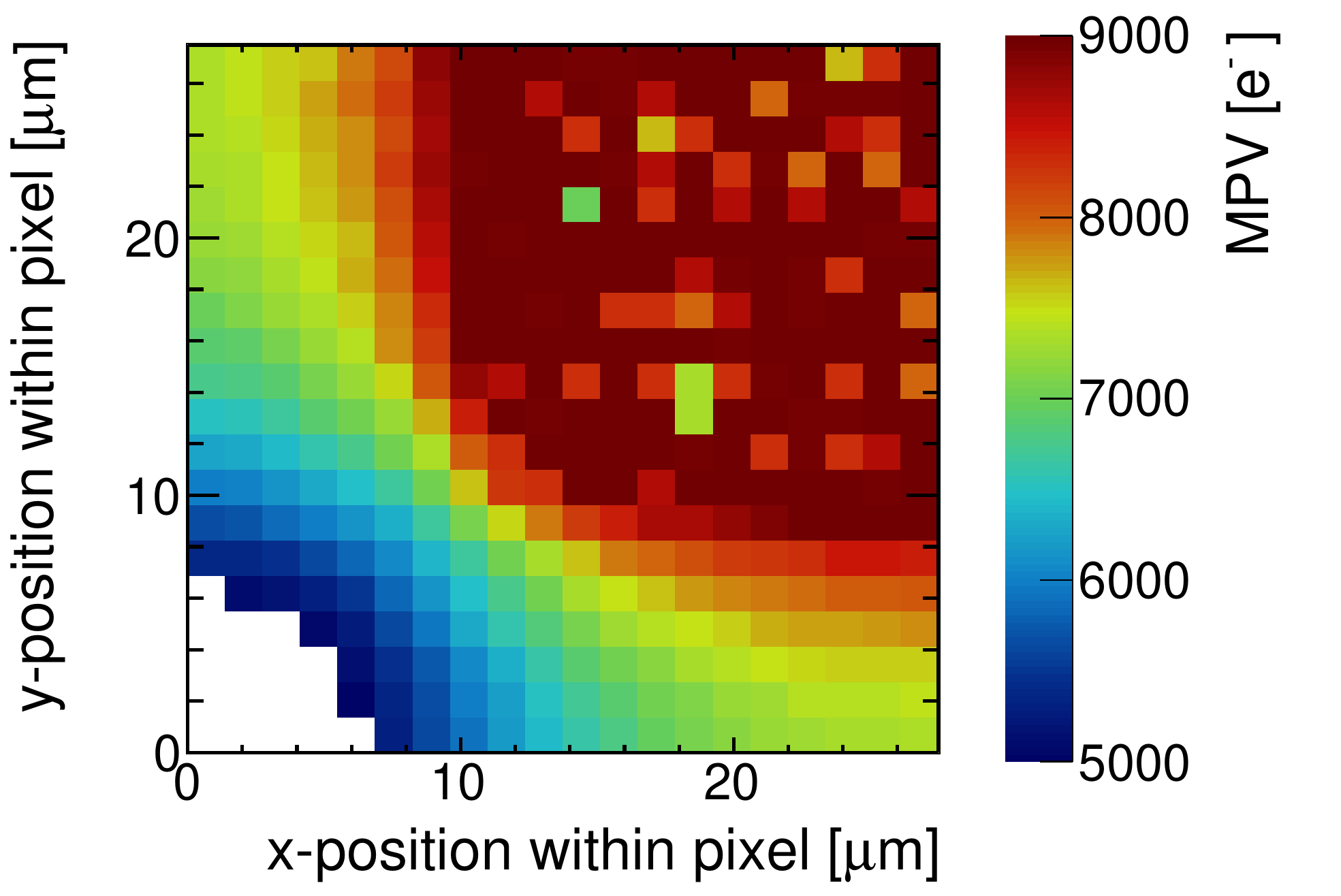}} \quad
  {\includegraphics[width=0.47\textwidth]{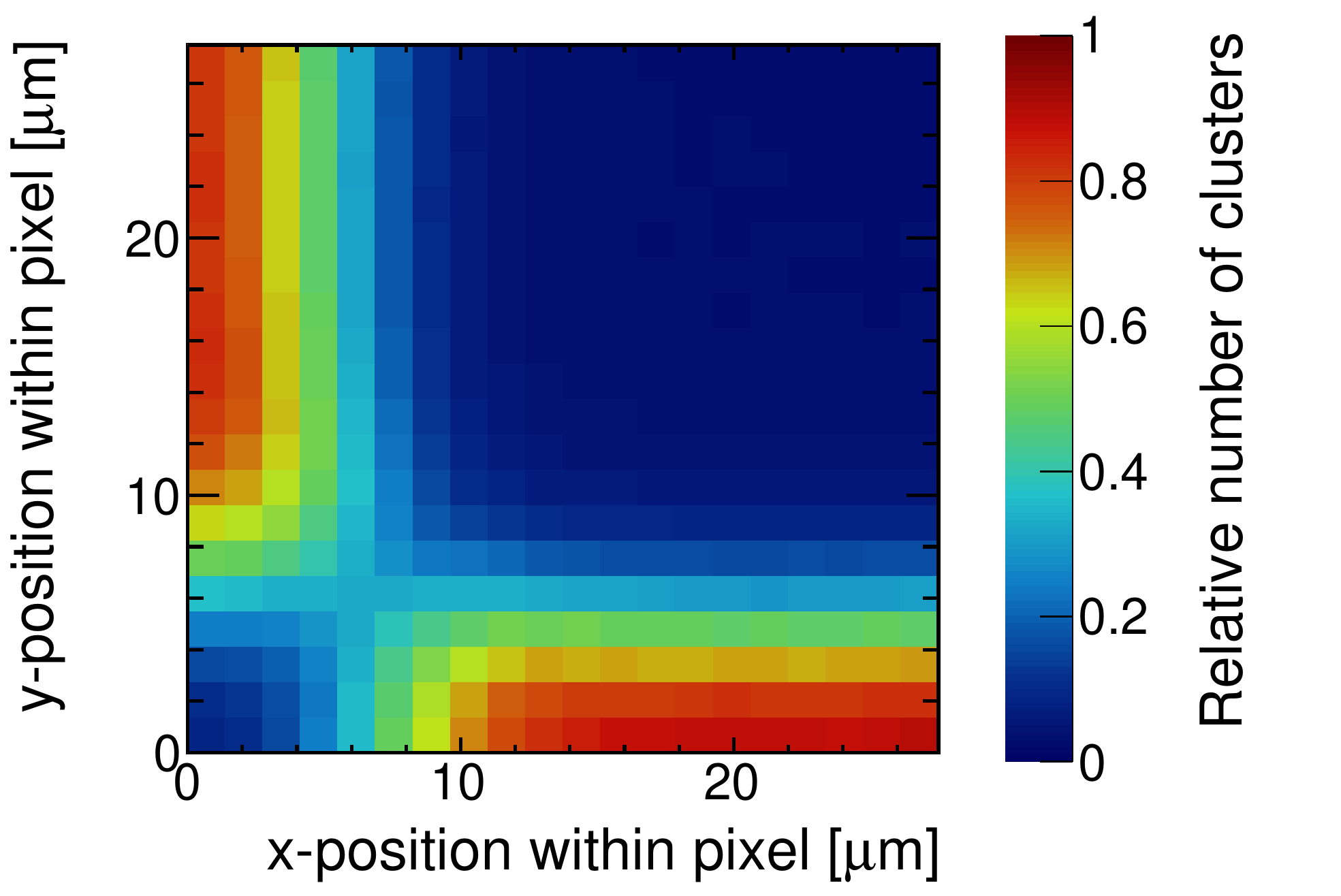}} \\
  {\includegraphics[width=0.47\textwidth]{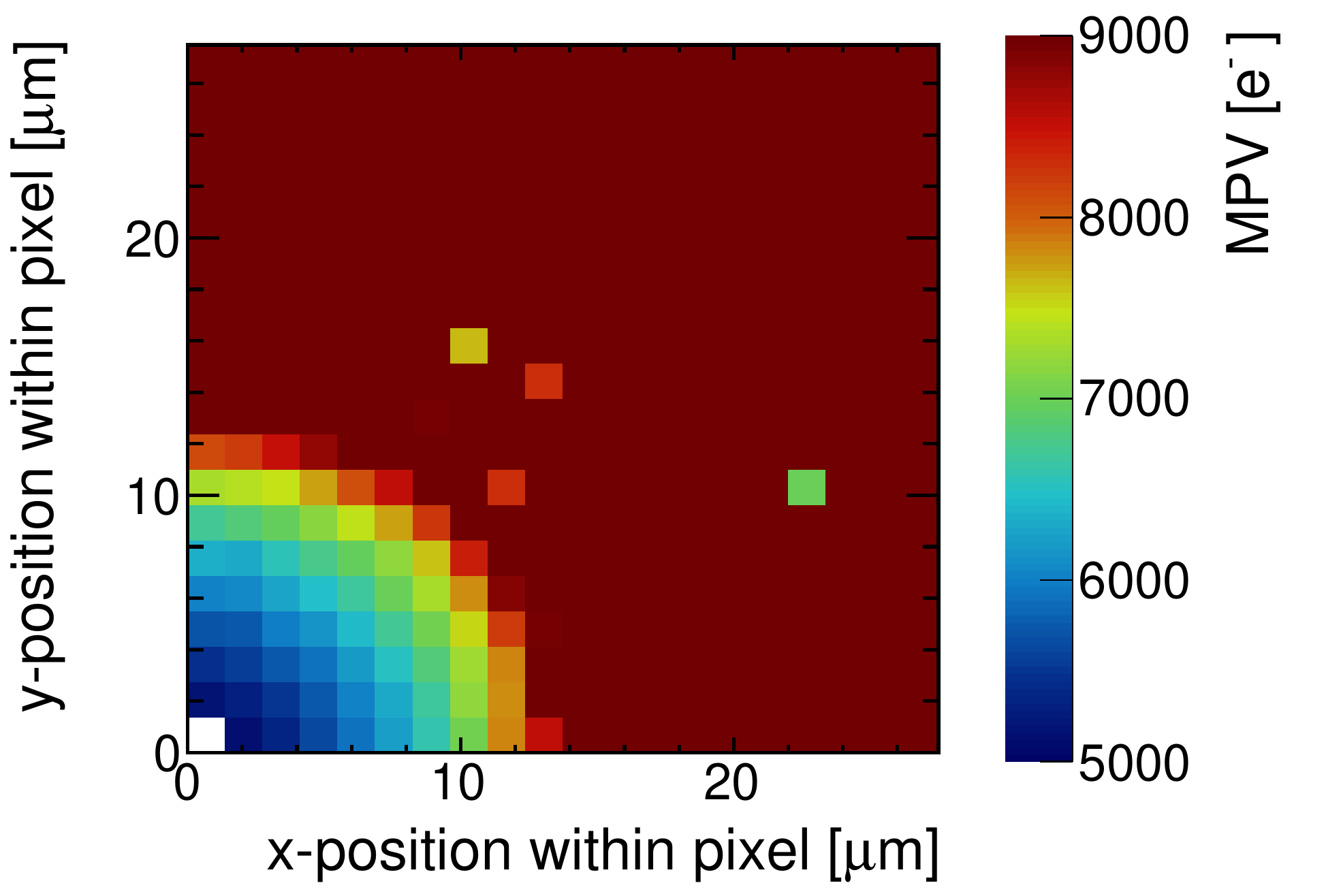}} \quad
  {\includegraphics[width=0.47\textwidth]{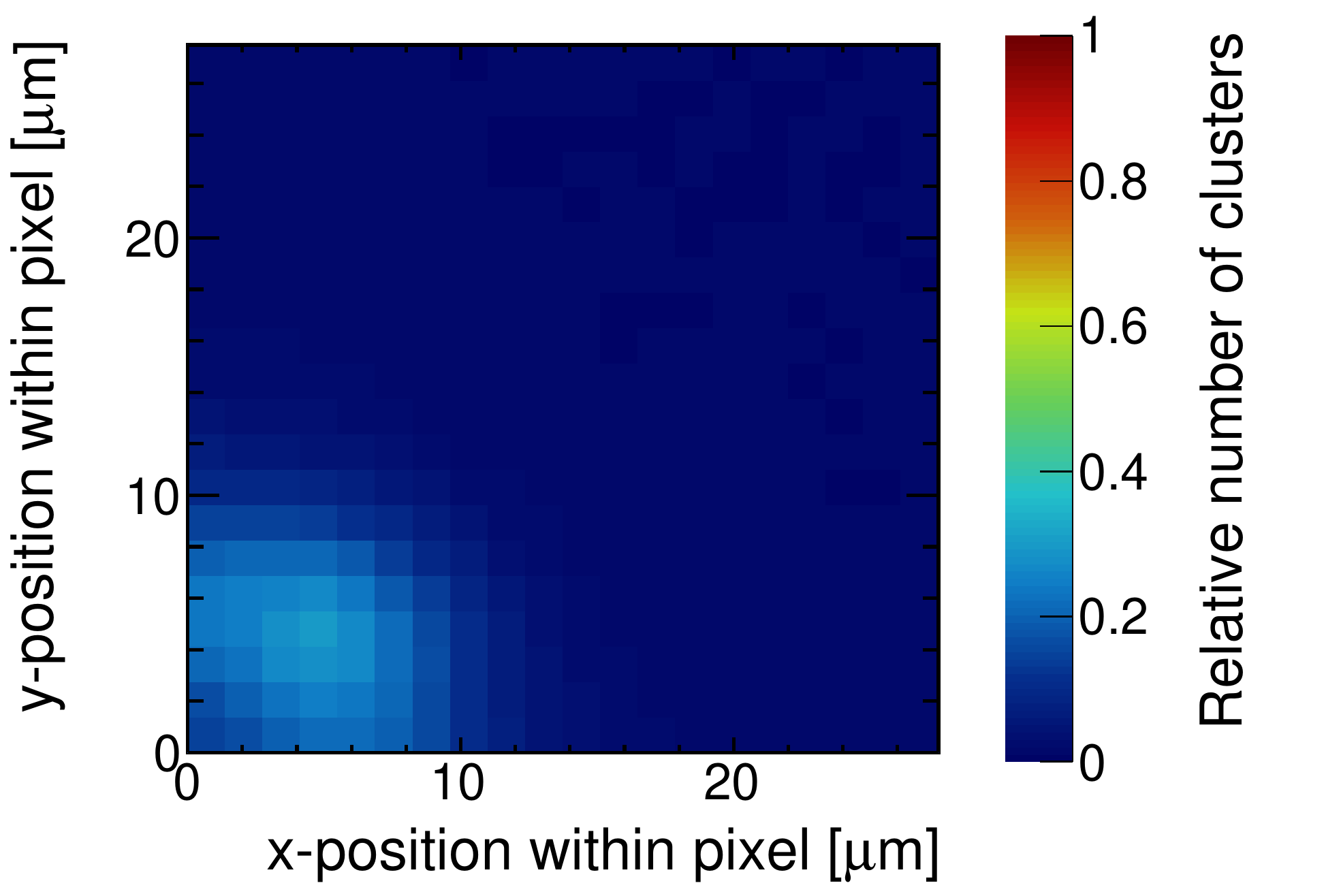}} 
  {\includegraphics[width=0.47\textwidth]{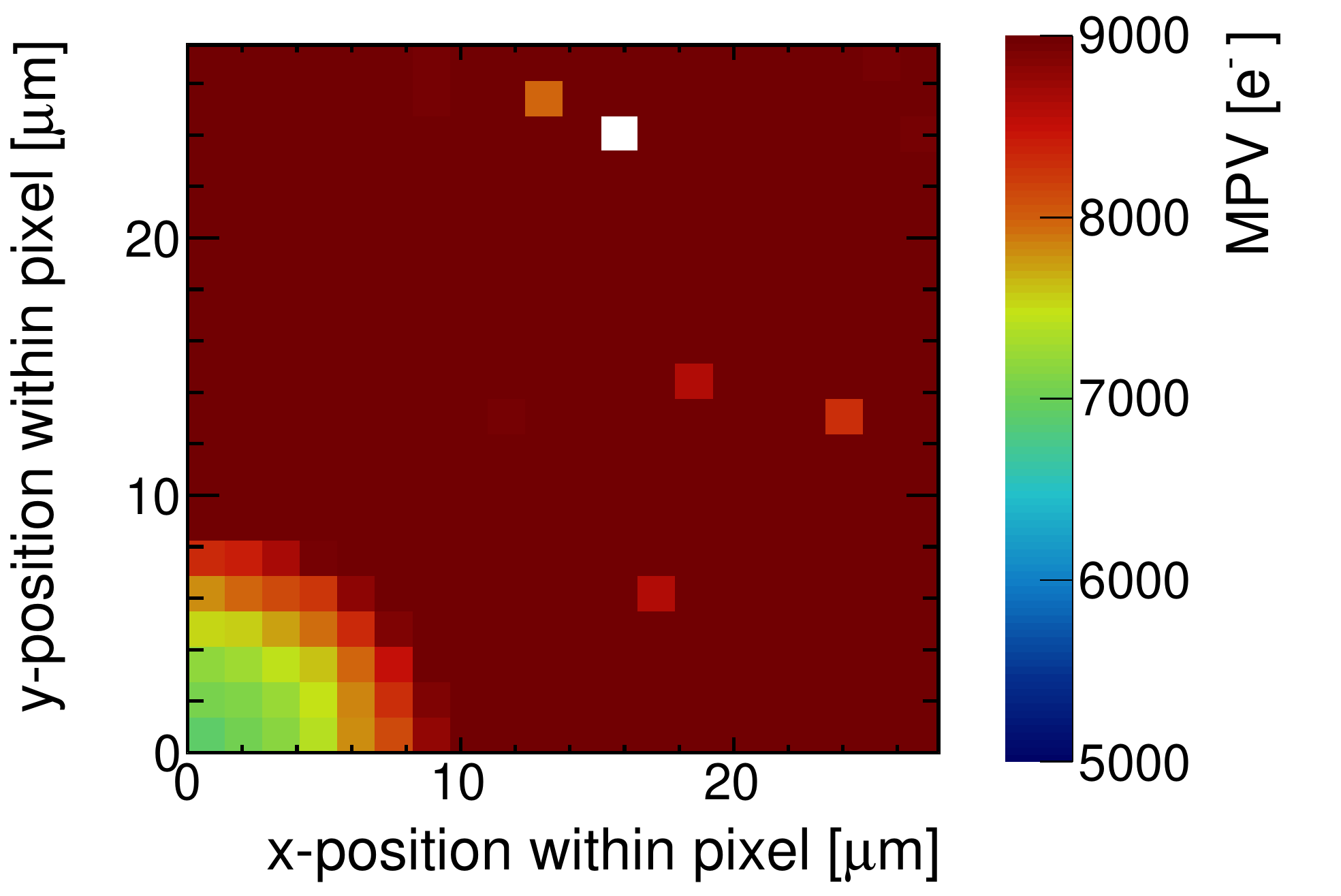}} \quad
  {\includegraphics[width=0.47\textwidth]{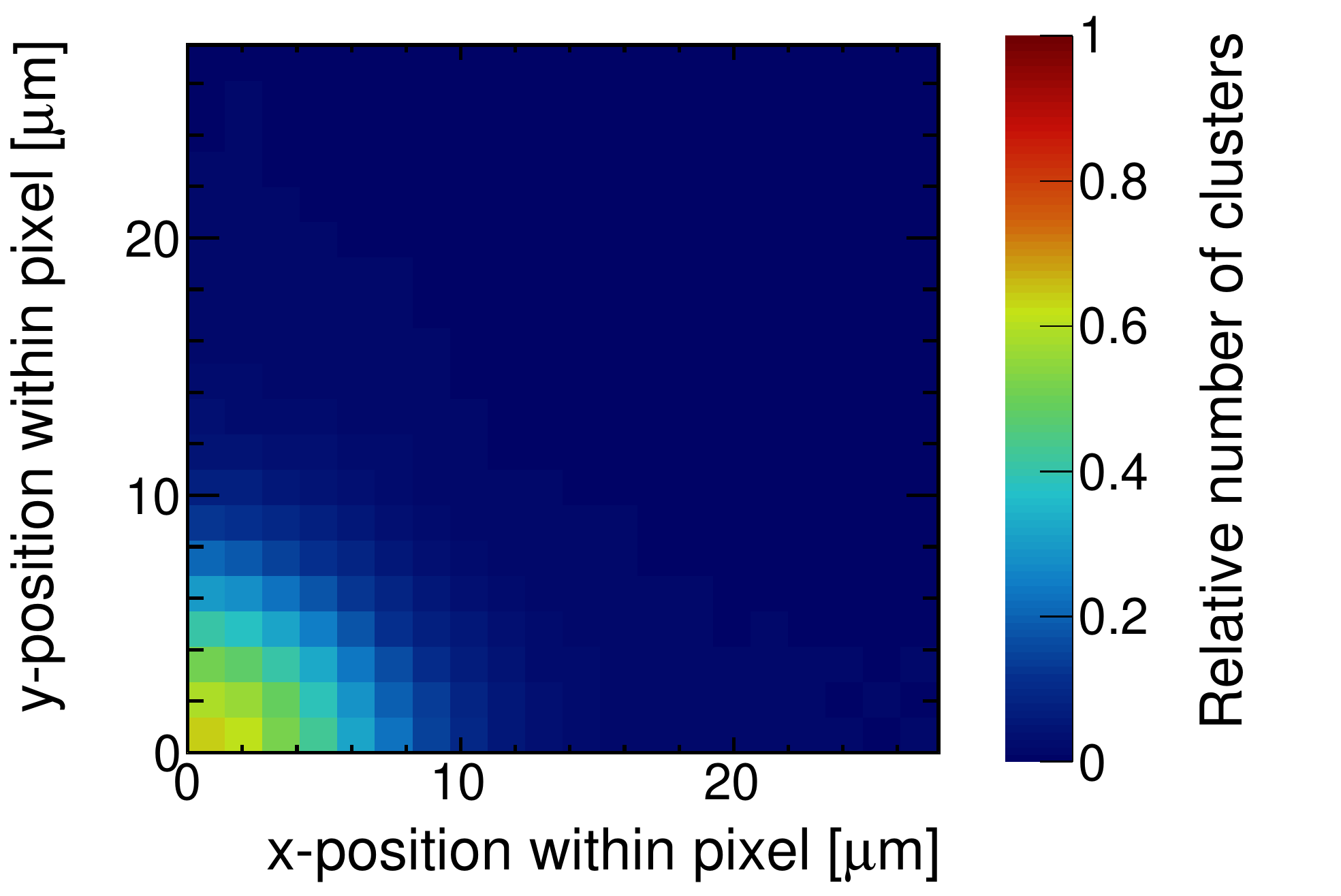}} 
  \caption{MPV for different intrapixel clusters positions (left) and the relative number of clusters (right) for cluster size one (first row), two (second row), three (third row), and four (last row). )}
  \label{fig:AppendixDifferentSizes}
\end{figure}

The MPV distribution as a function of the intrapixel position is split by the size of the cluster in order to study the origin of this structure, as is shown in \fig\ref{fig:AppendixDifferentSizes}, where the fraction of clusters of each size is also indicated. 
From the intrapixel position of the different cluster sizes it can be seen that the contribution to the band is mainly from clusters of size two or three, while the contribution to the corner is mainly from four-hit clusters. 
Because for four hit clusters all three neighbouring pixels must also measure a hit, there is no possibility to have undetected charge due to the signal in the neighbouring pixels being below the detection threshold. 
Conversely, for the other cluster sizes, there is the possibility that a neighbouring pixel collected some of the charge, but that it did not pass the detection threshold. 
If this occurs, the measured cluster charge is incomplete. 
For a single pixel cluster up to three contributions could be missing.
This inefficiency is unlikely to occur for hits in the middle of the pixel due to the limited size of the charge cloud, while it is more likely to happen to hits near the corners. 

\clearpage
\addcontentsline{toc}{section}{References}
\setboolean{inbibliography}{true}
\bibliographystyle{JHEP}
\bibliography{main}

\end{document}